\documentclass[11pt,prd,reprint,twocolumn,notitlepage,nofootinbib]{revtex4-1}

\usepackage{amsmath}
\usepackage{amsfonts}
\usepackage{graphicx}
\usepackage{aas_macros}
\usepackage{color}

\begin{document}

\title{New advances in the Gaussian-process approach to pulsar-timing data analysis}
\author{Rutger van Haasteren, Michele Vallisneri}
\affiliation{Jet Propulsion Laboratory, California Institute of Technology, Pasadena CA 91109}

\date{June 13, 2014}

\begin{abstract}
    In this work we review the application of the theory of Gaussian processes
    to the modeling of noise in pulsar-timing data analysis,
    and we derive various useful and optimized representations for the likelihood
    expressions that are needed in Bayesian inference on pulsar-timing-array datasets.
    The resulting viewpoint and formalism lead us to two improved
    parameter-sampling schemes inspired by Gibbs sampling.
    The new schemes have vastly lower chain autocorrelation lengths than the
    Markov Chain Monte Carlo methods currently used in pulsar-timing data analysis,
	potentially speeding up Bayesian inference by orders of magnitude.
	The new schemes can be used for a full-noise-model analysis of the large datasets assembled
	by the International Pulsar Timing Array collaboration, which present a serious
	computational challenge to existing methods.
\end{abstract}

\maketitle

\section{Introduction} \label{sec:intro}
    The high-precision timing of the radio emission from pulsars has proved to be a valuable tool for probing a wide range of science. Besides great successes such as the first indirect confirmation of the emission of  gravitational waves (GWs, \cite{1982ApJ...253..908T}), and very accurate tests of general relativity \cite{2006Sci...314...97K}, pulsar timing is now used in projects that aim to \emph{directly} detect low-frequency GWs ($10^{-9}$--$10^{-8}$ Hz) from extra-Galactic sources by using a set of Galactic millisecond pulsars (MSPs) as nearly perfect Einstein clocks \citep{1990ApJ...361..300F}, thanks to the exceptional regularity of their pulses---once many physical effects, such as the pulsar local and motion relative to the Earth, its binary dynamics if it has a companion, the propagation of pulses through the interstellar medium, and the intrinsic evolution of pulsar spin, are modeled accurately (indeed, an accurate \emph{timing model} must account for every rotation of the pulsar across observation epochs). The presence of GWs affects the propagation of the pulses from the pulsar to the Earth, creating detectable deviations from the strict periodicity of the pulse times of arrival (TOAs) \citep{1975GReGr...6..439E,1978SvA....22...36S,1979ApJ...234.1100D}.

    In the last decade, scientists seeking to detect GWs with pulsar timing have organized in  Pulsar Timing Array (PTA) projects around the globe: the European Pulsar Timing Array (EPTA, \cite{2011MNRAS.414.3117V,Kramer2013}), the North American Nanohertz Observatory for Gravitational Waves (NANOGrav, \cite{2013ApJ...762...94D,McLaughlin2013}), and the Australian Parkes Pulsar Timing Array (\hbox{PPTA}, \cite{2013PASA...30...17M,Hobbs2013}), which have now joined into a global collaboration, the International Pulsar Timing Array (IPTA, \cite{2010CQGra..27h4013H,Manchester2013}).
	Each PTA has now collected regular observations of tens of MSPs across several years, creating datasets of ever-increasing sensitivity to low-frequency GWs.
    As a result, a significant amount of effort has already been placed into the development of sophisticated data-analysis methods to extract GWs from pulsar TOAs, both for stochastic GW background signals
    \cite[among others:][]{2005ApJ...625L.123J, 2006ApJ...653.1571J, 2009MNRAS.395.1005V, 2009PhRvD..79h4030A, 2013PhRvD..87j4021L, 2013arXiv1310.4569S, 2013ApJ...762...94D, 2013PhRvD..88h4001T}, and continuous waves
    \cite[for instance:][]{2010MNRAS.407..669Y, 2010arXiv1008.1782C, 2010PhRvD..81j4008S, 2010MNRAS.401.2372V, 2011MNRAS.414.3251L, 2012ApJ...756..175E, 2012PhRvD..85d4034B, 2013PhRvD..87f4036P}.
    Many such methods, and especially those based on Bayesian principles, are very computationally intensive, and therefore slow. Although work is ongoing on their acceleration, large datasets such as those integrated by the International Pulsar Timing Array (IPTA) are still very challenging to analyze.
	
	Much of the sophistication required in PTA data analysis is concerned with the description of noise. GWs must be extracted from \emph{timing residuals} (the differences between the observed TOAs and the best timing-model fits), which include measurement errors but also other types of noise, such as ``red'' spin noise (or ``timing noise'', the long-term drifts in the rotational frequency of the pulsar), the time- and frequency-dependent delays due to pulse propagation through the interstellar medium, and effects that are correlated across pulsars, such as low-frequency drifts of atomic clocks, or inaccuracies in the Solar-system ephemerides. For a recent discussion of all of these, see \cite{2010arXiv1010.3785C, 2010ApJ...725.1607S}.
	Each of these noise sources must be distinguished from true GWs. The GWs themselves can have a stochastic character (as for the background from the superposition of signals from many supermassive black-hole binaries), in which case they can be extracted thanks to their correlations among pulsars.
	
	Modern data-analysis methods model the statistics of the noise components of timing residuals as \emph{time-correlated stochastic signals}, described by a power spectral density or a correlation function. This paper focuses on (and reviews) the description of stochastic signals as \emph{Gaussian processes}, the generalization of random variables to functions. This description was implicit in earlier contributions \cite[e.g.][]{2009MNRAS.395.1005V}, and we now make it fully explicit. Thus, we give a formal treatment of the Gaussian-process approach to pulsar-timing data analysis, and we derive (or rederive) various expressions, optimized in different ways, for the likelihood of the data in the presence of stochastic signals.
	We also describe and test two novel Bayesian sampling schemes, inspired by Gibbs sampling \cite{10.1109/TPAMI.1984.4767596}, which outperform the standard Markov-Chain Monte Carlo samplers used in pulsar-timing data analysis by greatly reducing the autocorrelation lengths of the chains.
	
    The outline of this paper is as follows. We introduce Gaussian processes in Sec.\ \ref{sec:GP}, and their application to pulsar-timing data in Sec.\ \ref{sec:GPPTA}. In Sec.\ \ref{sec:timingmodel} we discuss the analytical marginalization of likelihoods, and in Sec.\ \ref{sec:lowrank} we describe low-rank approximations of covariance matrices. Both techniques are crucial to high-performance analysis methods. In Sec.\ \ref{sec:sampling} we present our new-and-improved \emph{quasi-Gibbs} schemes, which we test on mock data in Sec.\ \ref{sec:simulations}. We end with our conclusions in Sec. \ref{sec:conclusions}.

\section{Gaussian processes} \label{sec:GP}

\emph{Gaussian processes} \cite{rasmussen2006gaussian} generalize the notion of Gaussian random variables to the case of an infinite number of degrees of freedom. They provide a modern treatment for \emph{process noise}, as defined in optimal filtering---a source of uncertainty distinct from measurement error, which represents unmodeled stochastic or systematic effects in the system under study.
More formally \cite{rasmussen2006gaussian}, a Gaussian process is a (possible infinite) ``collection of random variables, any finite number of which have a joint Gaussian distribution.'' This very property, which corresponds mathematically to the (always surprising) cancellations of chained exponential integrals, makes Gaussian processes especially suited to describing systems that have underlying continuous dynamics, yet are necessarily measured at a finite set of \emph{points} (which could be times, locations, or events).
Thanks to this property, the likelihood of a measured dataset as a function of the Gaussian-process parameters depends only on the behavior of the system at the points for which we have measurements; furthermore, it is especially convenient to interpolate or extrapolate inferences to points for which measurements were not made, or are not available.

A Gaussian process can be specified fully in one of two equivalent ways:
\begin{itemize}
\item As the sum $\sum_\mu \phi_\mu(x) w_\mu = \phi^T(x) w$ of a finite or infinite set $\{\phi_\mu(x)\}$ of deterministic basis functions, multiplied by the weights $w_\mu$, which are themselves Gaussian random variables with mean vector $w_\mu^0$ and covariance matrix $\Sigma_{\mu\nu}$. (This is the \emph{weight-space} view.)
\item As a continuous function $f(x)$, for which we prescribe the \emph{ensemble} mean $m(x) = \mathbb{E}[f(x)]$ and the covariance function $k(x,x') = \mathbb{E}[(f(x) - m(x))(f(x') - m(x'))]$. (This is the \emph{function-space} view.)
\end{itemize}
In the following we will adopt the simplifying but inessential assumption that $m(x) = w^0_\mu = 0$.
The duality between the two views and specifications is encapsulated by the covariance-function expansion
\begin{equation}
k(x,x') = \sum_{\mu,\nu} \phi_\mu(x) \Sigma_{\mu\nu} \phi_\nu(x').
\end{equation}
Indeed, Mercer's theorem \cite{rasmussen2006gaussian} ensures that a (possibly infinite) basis-function expansion exists for every positive-definite covariance $k(x,x')$. The power of switching between the dual views is manifest in the two equivalent expressions for the likelihood of a vector $y_i$ of observations of the Gaussian process, taken at the set of points $\{x_i\}$, and subject to Gaussian measurement noise $\epsilon_i$ with covariance matrix $N_{ij}$ \cite{rasmussen2006gaussian}:
\begin{widetext}
\begin{equation}
\begin{aligned}
p(y_i|w_\mu,\mathrm{GP}) &= \frac{\mathrm{e}^{-\frac{1}{2}\sum_{i,j}\bigl(y_i - \sum_\mu \phi_\mu(x_i) w_\mu \bigr) (N_{ij})^{-1} \bigl(y_j - \sum_\mu \phi_\mu(x_j) w_\nu \bigr)}}{\sqrt{(2 \pi)^n \det N}} \times
\frac{e^{-\frac{1}{2} \sum_{\mu\nu} w_\mu (\Sigma_{\mu\nu})^{-1} w_\nu}}{\sqrt{(2 \pi)^m \det \Sigma}} \\
p(y_i|\mathrm{GP}) &= \frac{\mathrm{e}^{-\frac{1}{2} \sum_{i,j} y_i \bigl(N_{ij} + K_{ij}\bigr)^{-1} y_j}}{\sqrt{(2 \pi)^n \det (N + K)}},
\quad \mathrm{with} \quad
K_{ij} = k(x_i,x_j) = \sum_{\mu\nu} \phi_\mu(x_i) \Sigma_{\mu \nu} \phi_\nu(x_j),
\end{aligned}
\label{eq:gplike}
\end{equation}
\end{widetext}
where $i,j = 1,\ldots,n$ and $\mu,\nu = 1,\ldots,m$, and $K_{ij}$ is the Gaussian-process \emph{covariance matrix} (i.e., the covariance function evaluated at the measured points). The first expression in Eq.\ \eqref{eq:gplike} shows the explicit dependence of the likelihood on the basis-function weights; the second, which is obtained by integrating over the $w^\mu$, is in effect the \emph{marginal likelihood} of the data given the Gaussian-process hypothesis, in a compact form that is especially useful if $k(x,x')$ [or equivalently the $\phi_\mu(x)$ and $\Sigma_{\mu\nu}$] are taken to be functions of a vector of \emph{hyperparameters}, such as the spectral amplitude and slope for power-law noise.

We take a moment to restate this important result: compared to the full likelihood $p(y_i|w_\mu,\mathrm{GP})$, the marginalized likelihood $p(y_i|\mathrm{GP})$ has been \emph{integrated with respect to all possible values of the Gaussian process at the measured points and everywhere else}, subject to the probabilistic constraints given by the noisy measurements. In a Bayesian framework, $p(y_i|\mathrm{GP})$ leads directly to the posterior probability for the hyperparameters.
If, conversely, we are interested in the inferred values of the Gaussian process \emph{given the observations}, it can be shown \cite{rasmussen2006gaussian} that at any points $x'$ and $x''$ (whether observed or not) the process is normally distributed with mean $\bar{x}' = \sum_{i,j} k(x',x_i)(N_{ij} + K_{ij})^{-1} y_j$ (and likewise for $x''$) and covariance $C(x',x'') = k(x',x'') - \sum_{ij} k(x',x_i)(N_{ij} + K_{ij})^{-1} k(x_j,x'')$. This equality has been re-derived and used in various forms in pulsar timing, for example for analytical marginalisation of the timing model parameters \cite{2009MNRAS.395.1005V}, and for reconstruction of DM variations \cite{2014MNRAS.441.2831L}.

For later reference, we rewrite Eq.\ \eqref{eq:gplike} using a looser notation where we omit vector indices, replace summations by vector--matrix multiplications, work with log-likelihoods, and adopt a special notation for normal-distribution normalization constants:
\begin{widetext}
\begin{equation}
\begin{aligned}
\log p(y|w,\mathrm{GP}) =& -\frac{1}{2} \bigl(y - \Phi^T(x) w\bigr)^T N^{-1} \bigl(y - \Phi^T(x) w\bigr) - \frac{1}{2} w^T \Sigma^{-1} w - \log \mathcal{N}_{n,N} - \log \mathcal{N}_{m,\Sigma}, \\
\log p(y|\mathrm{GP}) =& -\frac{1}{2} y^T (N + K)^{-1} y - \log \mathcal{N}_{n,N+K},
\end{aligned}
\label{eq:loggplike}
\end{equation}
where $n$ and $m$ are the sizes of the squares matrices $N$ and $\Sigma$, and where
\begin{equation}
\quad \Phi_{\mu,i} = \phi_\mu(x_i), \quad K = \Phi^T \Sigma \Phi, \quad \mathrm{and} \quad \log \mathcal{N}_{p,X} = \frac{p}{2} \log (2\pi) + \frac{1}{2} \log \det X.
\end{equation}
\end{widetext}
Gaussian processes have been studied for a long time in statistics; in the last 20 years they have received renewed attention in the fields of \emph{machine learning} and statistical inference \cite{rasmussen2006gaussian}.

\section{The Gaussian-process approach to pulsar-timing noise} \label{sec:GPPTA}

% We shall first review the task of pulsar timing, while establishing some dictionary.
In pulsar timing, the properties of the emitting system are inferred from the repeated timing of the its pulses. For millisecond pulsars, a large number of pulses is collected during each \emph{epoch} of observation. Each single \emph{time of arrival} (TOA) is determined by \emph{folding} the pulses with respect to fiducial period, and by cross-correlating the folded \emph{profile} to an independently determined \emph{template}; this process produces also an estimated measurement uncertainty (known as \emph{radiometer noise}) for each TOA \cite{2004hpa..book.....L}, which can be understood qualitatively as the width of the cross-correlation pattern around its maximum.
The TOAs can be predicted deterministically using models that include the
astrometric and physical parameters of the source (such as its sky position and
proper motion) and the intrinsic evolution of the pulsar spin frequency, as well as binary-orbit parameters for pulsars with a companion. Fitting a deterministic TOA model to a set of observed TOAs results in a \emph{timing solution}. The differences between observed and modeled TOAs are known as \emph{residuals}, and the best-fit model is usually chosen as the one that minimizes the root-mean-square uncertainty-weighted residual.

The gist of the Gaussian-process approach to inferring the noise properties of timing datasets and to searching for GW imprints in the TOAs is this:
\emph{the set of best-fit residuals for one or more pulsars is modeled as a sum of Gaussian processes}, which may include:
\begin{enumerate}
\item effects due to the necessarily imperfect determination of the timing solution;
\item additional observational errors not included in the cross-correlation estimate of TOA uncertainties;
\item sources of time-correlated or uncorrelated noise intrinsic to the pulsar;
\item effects due to the propagation of the pulses through the interstellar medium;
\item \emph{common-mode} effects that are correlated among multiple pulsars, such as those due to the presence of stochastic GWs or reference clock errors. 
\end{enumerate}
In this approach, we specify the covariance function or matrix for each Gaussian-process component (except for timing-solution errors, which are easiest to specify using basis functions) as functions of a set of hyperparameters, and we deploy the machinery of Bayesian inference to derive posterior distributions for the hyperparameters (and to characterize or marginalize over the timing-solution errors).
The approach was first formulated by van Haasteren and Levin \cite{2009MNRAS.395.1005V,2013MNRAS.428.1147V}, without drawing an explicit link to the theory of Gaussian processes, and in effect rederiving basic results such as Eq.\ \eqref{eq:loggplike} as probability manipulations in Bayesian inference.

Mathematically, we write the residuals $y$ as the sum
\begin{equation}
y(\theta) = \sum_{(A)} y^{(A)}(\theta^{(A)}) + \epsilon,
\label{eq:residuals}
\end{equation}
where the vector $\epsilon$ denotes measurement errors (which are taken to be Gaussian with covariance matrix $N$); where the set $\{y^{(A)}(\theta^{(A)})\}$ includes one or more of the Gaussian processes discussed above, with $\theta^{(A)}$ the hyperparameters appropriate for each; and where $\theta$ denotes the collection of all $\theta^{(A)}$. The crucial result from Gaussian-process theory, which enables Bayesian inference on the $\theta$, is the fact that the marginal likelihood $p(y|\theta, \mathrm{GP})$ can be written simply as
\begin{equation}
\begin{aligned}
\log p(y|\theta, \mathrm{GP}) = &
-\frac{1}{2} y^T (N + \sum_{(A)} K^{(A)})^{-1} y \\
& - \log \mathcal{N}_{n,N+\sum_{(A)} K^{(A)}}.
\label{eq:fulllike}
\end{aligned}
\end{equation}
(In fact, this simple description requires two slight complications: first, the timing-solution errors are usually given a special treatment, discussed in Secs.\ \ref{subsec:timingerrors} and \ref{sec:timingmodel}; second, the measurement-error matrix $N$ is also parameterized by one or more hyperparameters, as described in Sec.\ \ref{subsec:equadjitter} below.)

For one choice of hyperparameters, and under the assumption that $K$ is a dense matrix, the task of evaluating a likelihood for a dataset of $n$ TOAs using Eq.\ \eqref{eq:fulllike} involves the $O(n^2)$ computation of the total covariance matrix $N + K = N + \sum_{(A)} K^{(A)}$, the $O(n^3)$ computation of its determinant and inverse,\footnote{The best known algorithms have slightly lower exponents, but they are not always available in practical computational setups.} and the $O(n^2 + n)$ multiplication of the inverse covariance into the $y$. The cost of the inverse usually dominates the accounting.
Instead of computing $(N + K)^{-1}$ explicitly, one may obtain the upper-triangular decomposition $N + K = U^*U$, which yields the determinant as the product of the squared diagonal elements, and then compute $y^T (N+K)^{-1} y$ as $y^T (U \backslash (U^T \backslash y))$, where we used the MATLAB notation $A \backslash b$ for the solution $x$ of $A x = b$. The decomposition is again $O(n^3)$, but with a smaller numerical constant, while the linear-system solutions are $O(n^2)$.

Although the individual covariance matrices in the sum $\sum_{(A)} K^{(A)}$ are positive definite by the very definition of covariance, they may have very high condition numbers \cite{golub2012matrix} and thus they may be difficult to invert (or decompose) numerically. Nevertheless, the inversion of $N + K$ is usually \emph{regularized} by the measurement-error matrix $N$, which is typically diagonal, with elements that are large compared to the $K^{(A)}$. In the course of Bayesian inference, one may yet encounter corners of (hyper-)parameter space where $N + K$ becomes numerically singular; it has been our practice to assign a likelihood of 0 to those locations.

We now examine the individual Gaussian-process components of pulsar-timing models, and discuss the forms of covariance matrices appropriate for each.

\subsection{Timing-solution errors}
\label{subsec:timingerrors}

The best-fit timing solution for a set of TOAs is typically derived under the
assumption that the template--profile alignment uncertainties due to radiometer noise are the only source of
noise.\footnote{It is however becoming increasingly common to adopt more sophisticated noise models in timing work, following Refs.\ \cite{2011MNRAS.418..561C,2013MNRAS.428.1147V,2014MNRAS.437.3004L,2014MNRAS.440.1446V}.}
Even in that case, the resulting timing-model parameters would be slightly wrong because the minimum-residual solution always overfits the noise; in reality, the best-fit parameters will be systematically biased by the other unmodeled sources of noise.

If however the best-fit solution is sufficiently close to the truth and the various noise components are not too overwhelming, the component of the residuals due to timing-solution errors may be expressed as
\begin{equation}
y^{(\mathrm{TS})} = \sum_a \phi_a^{(\mathrm{TS})}(t) \delta \eta_a \equiv M \delta \eta,
\end{equation}
where $\delta \eta_a$ is the $p$-dimensional vector of the parameter errors $\delta \eta_a \equiv \eta_a^\text{best-fit} - \eta_a^\mathrm{true}$, where the $\phi_a^{(\mathrm{TS})}(t)$ are the partial derivatives of the TOAs with respect to the $\eta_a$, evaluated at $\eta^\text{best-fit}$, and where $M$ is the \emph{design matrix}\footnote{The design matrix yields the least-squares timing solution as the endpoint of the iteration $M(\eta^{[i]}) \Delta \eta^{[i+1]} = \mathrm{TOA}^\mathrm{obs} - \mathrm{TOA}^\mathrm{model}(\eta^{[i]})$, $\eta^{[i+1]} = \eta^{[i]} + \Delta \eta^{[i+1]}$.} $M_{ia} = \phi_a^{(\mathrm{TS})}(t_i)$.
The assumption that this linear regime for the $y^{(\mathrm{TS})}$ is actually realized in the course of Bayesian inference can be checked by carrying along the full nonlinear timing model, and exploring timing-model parameter space alongside with the Gaussian-process hyperparameters \cite{2014MNRAS.437.3004L,2014MNRAS.440.1446V}.

We do not usually deal explicitly with the covariance matrix that ensues from the basis functions $\phi_a^{(\mathrm{TS})}$, because it is awkward to attribute a prior covariance $\Sigma^\mathrm{TS}$ to the $\delta \eta_a$. Instead, we shall see in Sec.\ \ref{sec:timingmodel} how we can marginalize the likelihood with respect to an \emph{improper} prior for $\delta \eta$, which is equivalent to taking the limit $\lambda \rightarrow \infty$ for a prior of the form $\Sigma^\mathrm{TS} = \lambda I_p$.

\subsection{Measurement errors (EFAC, EQUAD, and jitter-like noise)}
\label{subsec:equadjitter}

We know empirically that the cross-correlation estimate of radiometer noise is not always correct; a common fix has been the inclusion in the model of a variable noise multiplier, known as EFAC. In fact, the physics of the measurement suggests that separate EFACs should be used for every receiver or backend represented in the dataset.
We know also that there are potential sources of measurement errors that are unrelated to radiometer noise; these have been represented as a white-noise component that adds to radiometer noise in quadrature, with an amplitude parameter known as EQUAD.\footnote{In the conventions of some timing packages, such as \texttt{Tempo2} \cite{tempo2web}, the EFAC parameter appears also in front of the EQUAD amplitude. We prefer to keep the two separate, since we believe that these hyperparameters should be uncorrelated.} Again, different EQUADs may be assigned to multiple receivers and backends.

Last, the circumstance that certain datasets (notably those collected by the NANOGrav collaboration \cite{2013ApJ...762...94D}) include TOAs measured at the very same time and for the very same set of folded pulses, but in neighboring frequency bands, creates the possibility of noise that is largely or entirely correlated among TOAs measured simultaneously, but entirely uncorrelated among TOAs taken at different times. Some, but perhaps not all, of this noise may be understood as \emph{pulse phase jitter} \cite{2010arXiv1010.3785C} caused by variable emission within pulsar magnetospheres.
% [MV: say more about estimating jitter amplitude at different frequencies? \rvh{Perhaps new paper by Shannon et al on jitter will be out in few days.}]

In the case of a single receiver/backend, the total covariance matrix for these three noise components can be written as
\begin{equation}
K^{(\mathrm{MN})} = E^2 \, n_i \delta_{ij} + Q^2 \, \delta_{ij} + J^2 \, \delta_{e(i)e(j)},
\label{eq:measnoise}
\end{equation}
where the indices $i$ and $j$ range over the TOAs; where the $n_i$ are the cross-correlation estimates of radiometer noise for each; where the $\delta$ are Kronecker deltas; and where $e(i)$ indexes the \emph{epochs} (i.e., reference measurement times) of each TOA. If the TOAs are sorted by epoch, the matrix $\delta_{e(i)e(j)}$ is block-diagonal, with each block consisting entirely of ones. Such a matrix has low rank corresponding to the number of epochs, which allows useful computational optimizations, discussed below in Sec.\ \ref{sec:lowrank}. 

It is largely a matter of taste (and sometimes, as we will see below, computational convenience) whether to include all three components in the notional measurement noise $\epsilon$ (in which case $N = K^{(\mathrm{MN})}$), or to designate EQUAD noise and jitter-like noise as separate Gaussian processes (in which case $K^{(\mathrm{MN})} = N + K^{(\mathrm{Q})} + K^{(\mathrm{J})}$).
For the case of multiple receivers and backends, separate EFAC, EQUAD, and jitter-like terms for each would appear in Eq.\ \eqref{eq:measnoise}, with each set of terms applying to a disjoint subset of TOAs. If the TOAs are sorted by receiver/backend, the total covariance matrix is block diagonal, and each block has the form of Eq.\ \eqref{eq:measnoise} with different $E$, $Q$, and $J$.

\subsection{Correlated pulsar noise}
\label{subsec:rednoise}

Millisecond pulsars are excellent clocks, but they are not perfect. Slight but measurable irregularities in their rotation (which may be due, for instance, to random angular-momentum exchanges between the normal and superfluid components of the pulsar \cite{2010arXiv1010.3785C}) create a time-correlated stochastic component in the TOAs that is referred to as ``timing noise'' or ``red spin noise.''
This component of timing residuals is typically modeled as a Gaussian, stationary random process, with power-law power spectral density:
\begin{equation}
P^{(\mathrm{PL})}(f) = A^{2} (f/\mathrm{yr}^{-1})^{-\gamma} \, \mathrm{yr}^{3},
\label{eq:powerlaw}
\end{equation}
where $f$ is the frequency, $A$ is a dimensionless amplitude and $\gamma$ is the spectral index of the power law (the alternative parametrization $\alpha = 3/2-\gamma/2$ is also in use). By way of the Wiener--Khinchin theorem,\footnote{For a stationary process for which $k(x',x'') = C(x' - x'') = C(\Delta x)$, the Wiener--Khinchin theorem relates the power spectral density $P(f)$ to the correlation function $C(\Delta t)$ by way of $C(\Delta t) = \int_0^\infty \cos(2 \pi f \Delta t)P(f)\mathrm{d}f$. The total variance of the process is then $C(0)$.} Eq.\ \eqref{eq:powerlaw} results in the correlation matrix
\begin{widetext}
\begin{equation}
    K^{(\mathrm{PL})}_{ij} = k^{(\mathrm{PL})}(t_i,t_j) = 
    A^{2} (f_L/\mathrm{yr}^{-1})^{1-\gamma}
	  \biggl\{ \Gamma(1-\gamma)\sin\Bigl(\frac{\pi\gamma}{2}\Bigr)
	  (f_{L}\tau_{ij})^{\gamma-1}
	-\sum_{n=0}^{\infty}
	  \frac{(-1)^{n} \, (f_{L}\tau_{ij})^{2n}}{(2n)!\,(2n+1-\gamma)}\biggr\},
    \label{eq:powerlawcovariance}
\end{equation}
\end{widetext}
where $\Gamma(\cdot)$ denotes the Euler gamma function, $\tau_{ij} = 2\pi|t_i-t_j|$
is the absolute difference of TOAs, and $f_L$ is a low-frequency cutoff that regularizes the Wiener--Khinchin integral.
The series in Eq.\ \eqref{eq:powerlawcovariance} sums up to ${}_1F_2(\{1/2-\gamma/2\},\{1/2,3/2-\gamma/2\},-(f_L \tau_{ij})^2/4) / (\gamma - 1)$, where ${}_1F_2$ is the generalized hypergeometric function given by \texttt{HypergeometricPFQ} in \textsl{Mathematica} and by \texttt{hyp1f2} in \texttt{scipy.special}.

Blandford and colleagues \cite{1984JApA....5..369B} and later other authors \cite{2009MNRAS.395.1005V,2012MNRAS.423.2642L,2013MNRAS.428.1147V} showed that the exact value of $f_L$ is irrelevant in pulsar applications, since it is absorbed in the fitting of the linear- and quadratic-spindowns term of the timing model, at least for $\gamma$ up to 7 (up to 5 using the linear term alone).
For $\gamma = 1$, the total variance $\int P(f) \mathrm{d}f$ becomes infinite even with the low-frequency cutoff.
Thus, the spectral index $\gamma$ is usually taken in the interval $[1,7]$, although imposing a high-frequency cutoff\footnote{This can be achieved by taking the difference of two expressions of the form \eqref{eq:powerlawcovariance} with different $f_L$.} makes it possible to reach $\gamma = 0$, which corresponds to band-limited white noise.

The evaluation of Eq.\ \eqref{eq:powerlawcovariance}
% \rvh{This is strange. Why does powerlawcovariance point to the wrong equation?} MV: there were two identical \labels. Fixed now.
is numerically delicate, so special care and tricks are needed.\footnote{Equation \eqref{eq:powerlawcovariance} becomes singular for some values of $\gamma$, so special-case expressions are required. An alternative, more benign low-frequency regularization is to redefine $P(f) = A^2((f \mathrm{yr})^2 + (f_L \mathrm{yr})^2)^{-\gamma/2}$, which leads to a $C(\tau)$ expressed in terms of modified Bessel functions of the second kind.} Furthermore, $K^{(\mathrm{PL})}_{ij}$ is a dense, full-rank matrix, so its use in computing residual likelihoods incurs the full $O(n^3)$ cost of matrix inversion. For $\gamma \gtrsim 6$ the matrix $K^{(\mathrm{PL})}_{ij}$ gains a very large condition number [on the order of $(f_L \min \tau)^{-\gamma}$] so the inversion can also be numerically unstable, although it may be regularized by the fact that we invert $N + K$ rather than $K$, where $N$ is diagonal and has relatively large elements.

Both problems are solved by an alternative approach that models correlated timing noise as a sum over a set of Fourier modes \cite{2013PhRvD..87j4021L}:
\begin{equation}
\label{eq:sumovermodes}
y^\mathrm{(FM)}(t) = \sum_{k=1}^{q} a_k \cos (2\pi k \, x) + b_k \sin (2\pi k \, x), 
\end{equation}
where $x = (t - t_0) / T$, with $t_0$ and $T$ the beginning and end of the observation span, respectively.
From a Gaussian-process perspective, this amounts simply to specifying the basis functions $\phi_\mu$ instead of the covariance function, and solving for the weights $w_\mu$ (here we subsume the cosines and sines, and their coefficients, into a single vector of bases of dimension $2q$).
This approach offers the additional freedom of specifying the prior weight covariance $\Sigma^{(\mathrm{FM})}_{\mu \nu}$ as function of a set of hyperparameters. For instance, a diagonal $\Sigma^{(\mathrm{FM})}_{\mu \nu}$ specifying a set of variances $\rho_\mu$, each shared by the $\cos$ and $\sin$ modes of the same frequency $f_\mu$, can be used for a form of spectral estimation \cite{2013PhRvD..87j4021L} (which is not quite ``model independent,'' as it is called in Ref.\ \cite{2013PhRvD..87j4021L}, since a prior for the $\rho_\mu$ is still required).

The Fourier-sum approach can be seen also as a suboptimal spectral approximation of the time-domain power-law covariance, by way of the fundamental Gaussian-process duality relation: 
\begin{equation}
K^{(\mathrm{PL})}_{ij} = \sum_{\mu\nu} \phi^{(\mathrm{FM})}_\mu(t_i) \Sigma^{(\mathrm{FM})}_{\mu\nu} \phi^{(\mathrm{FM})}_\nu(t_i)
% \quad \text{with} \quad
\label{eq:fouriersum}
\end{equation}
with
\begin{equation}
\Sigma^{(\mathrm{FM})}_{\mu\nu} = P^{(\mathrm{PL})}(f_\mu) \Delta f\delta_{\mu\nu} = P^{(\mathrm{PL})}(f_\mu) \delta_{\mu\nu} / T.
\end{equation}
The approximation is suboptimal both because we usually sum over a small number of modes (so it is a \emph{low-rank} approximation of a full-rank matrix), and because the modes are not the true eigenfunctions of $K^{(\mathrm{PL})}_{ij}$. However, in practice Eq.\ \eqref{eq:fouriersum} can be very accurate (especially if additional, logarithmically spaced modes are added at low frequencies \cite{vanHaasteren2014}). It can also offer very significant computational savings, because the inverse of a matrix expression involving low-rank addends can be computed very efficiently. We discuss this optimization extensively in Sec.~\ref{sec:lowrank} below.

\subsection{Propagation through the interstellar medium}
\label{subsec:dm}

Pulsar radio signals travel across the electromagnetically dispersive interstellar medium, incurring a frequency-dependent, stochastic phase delay known as dispersion-measure (DM) noise \cite{1984Natur.307..527A,2010arXiv1010.3785C}, given by
\begin{equation}
y^{(\mathrm{DM})} = \left(4.15 \times 10^{-3} \, \mathrm{s}\right)
\left(\frac{\mathrm{DM}}{\mathrm{pc} \, \mathrm{cm}^{-3}}\right)
\left(\frac{\nu}{\mathrm{GHz}}\right)^{-2}
\end{equation}
for the delay of a pulse measured at frequency $\nu$ with respect to a (hypothetical) pulse at infinite frequency. The time-dependent quantity $\mathrm{DM}$ is the column density of free electrons along the (time-changing) line of sight from the pulsar to the radiotelescope. See Lee and colleagues \cite{2014MNRAS.441.2831L} for a discussion of previous work to characterize DM variations and their impact on pulsar-timing GW searches. In the analysis of pulsar-timing datasets that comprise observations at multiple frequencies, DM variations have been modeled with timing-model parameters that describe $\mathrm{DM}(t)$ as a piecewise constant \cite{2013ApJ...762...94D} or linear \cite{2013MNRAS.429.2161K} function.
% \rvh{is that all what Keith et al do?}. MV: pretty much, with some sophistication about choosing the points...
Alternatively, one can try to solve for DM variations from the multifrequency observations at each epoch, effectively generating a reduced infinite-frequency dataset \cite{2014MNRAS.441.2831L,2014arXiv1402.1672P}.

In the context of the Gaussian-process approach, DM noise can be modeled as a correlated Gaussian process, with an additional dependence on the frequency at which each TOA was determined \cite{2014MNRAS.437.3004L,2014MNRAS.441.2831L}. For DM variations characterized by the power-law power spectral density
\begin{equation}
P^{(\mathrm{DM})}(f) = A_\mathrm{DM}^{2} (f/\mathrm{yr}^{-1})^{-\gamma_\mathrm{DM}} \, \mathrm{yr}^{3},
\end{equation}
the \emph{timing-residual} covariance function is
\begin{multline}
K^{(\mathrm{DM})}_{ij} = 
k^{(\mathrm{DM})}(t_i,t_j) = \left(4.15 \times 10^{-3} \, \mathrm{s}\right)^2 \\
\times \left(\frac{\nu_i \nu_j}{\mathrm{GHz}}\right)^{-2}
%\left(\frac{\nu_j}{\mathrm{GHz}}\right)^{-2}
\left|k^{(\mathrm{PL})}(t_i,t_j)\right|_{A \rightarrow A_\mathrm{DM}, \gamma \rightarrow \gamma_\mathrm{DM}},
\label{eq:dmcovariance}
\end{multline}
where the last term is given by the red-noise power-law covariance Eq.\ \eqref{eq:powerlawcovariance} after replacing $A$ and $\gamma$ with their DM counterparts. For a Kolmogorov DM spectrum resulting from plasma turbulence, $\gamma_\mathrm{DM} = 11/3$ \cite{2007MNRAS.378..493Y,2013MNRAS.429.2161K}.

The caveats given above for $K^{(\mathrm{PL})}_{ij}$ apply also to the evaluation of $K^{(\mathrm{DM})}_{ij}$. It is also possible to model $y^{(\mathrm{DM})}$ as a sum over basis functions, in analogy to Eq.\ \eqref{eq:sumovermodes}, using either a ``spectral-estimation'' or power-law prior. If the basis functions are Fourier modes at multiples of the fundamental frequency $1/T$ (with $T$ the duration of the dataset), the very low-frequency behavior of the Gaussian process is not modeled well \cite{2014MNRAS.437.3004L}; this can be remedied by enhancing the timing-model design matrix with a term similar to quadratic spindown, but with $\nu^{-2}$ frequency dependence \cite{2014MNRAS.437.3004L}, or by adding more modes at low non-Fourier frequencies \cite{vanHaasteren2014}.

\subsection{Gravitational waves and clock errors}
\label{subsec:commonmode}

Pulsar TOAs carry an imprint of the space-time perturbations (i.e., GWs) that they traverse as they travel from their neutron-star source to the Earth \cite{1979ApJ...234.1100D}. For an individual source of plane GWs, the frequency-shifting \emph{Doppler response} of the pulsar-to-radiotelescope baseline includes an \emph{Earth term} proportional (times geometric factors) to the GW strain at the event (time and place) of pulse reception, and a \emph{pulsar term} proportional to the GW strain at the event of pulse emission \cite{1975GReGr...6..439E}. Integrating both terms yields the TOA response, modulo a constant time offset that is degenerate with the initial-phase parameter of the timing model (see, e.g., \cite{2004ApJ...606..799J} for the case of GWs from a black-hole binary). In the Gaussian-process approach to pulsar-timing analysis, such a deterministic signal would not be modeled as a stochastic process, but rather it would be subtracted from the residuals before evaluating their likelihood.

By contrast, a \emph{stochastic} background of GWs can be modeled as a Gaussian process and included in Eq.\ \eqref{eq:residuals}. Various commonly considered backgrounds have a power-law power spectral density: 
\begin{equation}
P^{(\mathrm{GW})}(f) = \frac{A_\mathrm{GW}^{2}}{12 \pi^2} (f/\mathrm{yr}^{-1})^{-\gamma_\mathrm{GW}} \, \mathrm{yr}^{3},
\label{eq:gwpowerlaw}
\end{equation}
where the $12 \pi^2$ factor follows from defining $A_\mathrm{GW}$ as the dimensionless characteristic strain $h_c$ at $f = 1/\mathrm{yr}$ \cite{2006ApJ...653.1571J},
\begin{equation}
h_c(f) = A_\mathrm{GW} (f/\mathrm{yr}^{-1})^{\alpha_\mathrm{GW}} \quad
\text{where} \quad \gamma_\mathrm{GW} = 3 - 2 \alpha_\mathrm{GW}. 
\end{equation}
%
% [MV: do we usually include the $1/12 \pi^2$ in the red-noise and DM definitions? \rvh{Yes, but only for convenience. I think we should make a point of not doing it for the DM variations}]
The spectral index $\gamma_\mathrm{GW}$ is $13/3$ (but possibly less at low frequencies) for the background from the sum of unresolved black-hole binaries \cite{2003ApJ...583..616J,2008MNRAS.390..192S,2013MNRAS.433L...1S}; $16/3$ for a background from cosmic superstrings \cite{2005PhRvD..71f3510D,2007PhRvL..98k1101S}; $5$ for a background of inflationary relics \cite{2005PhyU...48.1235G}. Nonstrictly power-law spectra are also possible, as in the case of the QCD phase transition \cite{2010PhRvD..82f3511C}.

Thus, a GW background can be modeled as a stochastic process with a time-domain covariance matrix analog to Eq.\ \eqref{eq:powerlawcovariance}, or with a Fourier-sum covariance analog to Eq.\ \eqref{eq:fouriersum}. However, the very concept of pulsar timing array depends on the fact that the TOA imprints of stochastic GWs are \emph{correlated} among different pulsars. For an isotropic background, the correlation between the GW-induced residuals $y^{(\mathrm{GW})}_{ia}$ (for pulsar $a$) and $y^{(\mathrm{GW})}_{jb}$ (for pulsar $b$) is given by
\begin{equation}
\label{eq:multicov}
K^{(\mathrm{GW})}_{ia\,jb} = \zeta(\gamma_{ab}) \, k^{(\mathrm{GW})}(t_{ia},t_{jb})
\end{equation}
where $\zeta(\gamma_{ab})$ is the Hellings--Downs coefficient \cite{1983ApJ...265L..39H,2008ApJ...685.1304L} for the angle $\gamma_{ab}$ between the pulsars:
\newcommand{\singamma}{\sin^2 \! \Big(\frac{\gamma}{2}\Big)}
\begin{equation}
\begin{aligned}
% \zeta(\gamma_{ab}) = \frac{3}{2} \frac{1 - \cos \gamma_{ab}}{2} \log \left(\frac{1 - \cos \gamma_{ab}}{2}\right) - \frac{1}{4} \frac{1 - \cos \gamma_{ab}}{2} + \frac{1}{2} \quad
\zeta(\gamma_{ab}) & = \frac{3}{2} \singamma \log \singamma - \frac{1}{4} \singamma + \frac{1}{2},
% \quad \text{or} \quad
\\ \zeta(0) & = 1.
\label{eq:hellingsdowns}
\end{aligned}
\end{equation}
The correlations have a more complicated structure if the GW polarizations are not the two quadrupolar modes predicted by general relativity
\cite{2008ApJ...685.1304L,2011PhRvD..83l3529A,2012PhRvD..85h2001C}
or if the background is not isotropic
\cite{2013PhRvD..88h4001T,2013PhRvD..88f2005M}.

It follows that the full GW-background covariance matrix for a pulsar-timing-array dataset can be very large ($N \times N$, where $N = \sum_a n_a$ is the sum of the TOA counts for the individual pulsars); it is also dense, so its inversion is a computationally expensive proposition.
In Sec.~\ref{sec:fourier} we will see that modeling the GW background as a Fourier sum (and matching Fourier frequencies among pulsars) offers a useful shortcut.

The fact that correlations between a multitude of pulsars are used as a detection mechanism makes pulsar timing arrays robust detectors for gravitational waves. Noise can generally be expected to be uncorrelated between pulsars, and even without doing proper parameter estimation and noise analysis, a stochastic GW background can still be detected when enough pulsars are observed (Jenet et al 2005, Siemens et al 2013). However, GWs are not the only types of signals that can induce correlations. Slow drifts of atomic clocks can introduce an slight error in terrestrial time standards, which would manifest themselves as a common low-frequency signal in the signals of all pulsars (Hobbs et al. 2012). Such a correlated signal would be a source of noise when detecting a GW background, and it must be modelled appropriately. The clock signal consists of a time-correlated stochastic signal that is common to all pulsars. As such, we use the same models as for a GW background, except that we take $\zeta = 1$ instead of Eq.\ \eqref{eq:hellingsdowns}. Although the covariance matrix component of the clock signal is actually singular, in all realistic scenarios this is always regularized by the other constituents of the covariance matrix. If no regularising signal is present in the model as could be the case with mock data, it is trivial to replace it with a rank-reduced expansion similar to what we did in the previous sections.

Besides clock errors, another possible source of noise one could imagine are inaccuracies in the solar system ephemeris. Although these are unlikely to be a significant source of noise at the level of stochastic searches, if necessary they are easily modelled as a correlated stochastic signal. In that case, one should replace Eq.\ \eqref{eq:hellingsdowns} with $\zeta(\gamma_{ab})=\cos(\gamma_{ab})$.

\section{Marginalizing over timing-solution errors}
\label{sec:timingmodel}

As mentioned above, the timing-solution parameter errors $\delta \eta_a$ are usually given a special treatment: we can include them among the inferred parameters in a Bayesian analysis (i.e., among the parameters that would be sampled explicitly in a Markov-chain Monte Carlo run), and use a likelihood in the form
\begin{widetext}
\begin{equation}
\log p(y|\theta^\text{(non-TS)},\delta \eta_a) = -\frac{1}{2} (y - M \delta \eta)^T (N + K^{(\mathrm{non-TS})})^{-1} (y - M \delta \eta) - \log \mathcal{N}_{n,N + K^{(\mathrm{non-TS})}},
\label{eq:expliciteta}
\end{equation}
\end{widetext}
where the $\theta^\text{(non-TS)}$ denote all the model parameters other than the timing-model errors $\delta \eta_a$,
% \rvh{is $\theta^\text{(non-TS)}$ properly defined somewhere? Are these always hyperparameters?}
or we can treat them nonlinearly, as in Refs.\ \cite{2014MNRAS.440.1446V,2014MNRAS.437.3004L}, so that the residuals $y$ are recomputed from the full timing model for each value of the $\eta$ that we sample. This latter approach is desirable if we think that the functional dependence of the residuals on the $\eta$ may be significantly nonlinear within the relevant parameter ranges.

Otherwise, Eq.\ \eqref{eq:expliciteta} can be marginalized analytically over the $\delta \eta_a$ by computing the integral $\int p(y|\theta^\text{(non-TS)},\delta \eta_a) \mathrm{d}(\delta \eta_a)$. When doing so, we are in effect assuming an improper (infinitely vague) prior for the $\delta \eta_a$, which is acceptable from a Bayesian perspective as long as the observed data is informative with respect to those parameters.
The first authors to propose this marginalization were van Haasteren and Levin \cite{2009MNRAS.395.1005V}, who showed that
\begin{widetext}
\begin{equation}
\begin{aligned}
p(y|\theta^\text{(non-TS)}) &=
% \int p(y|\theta^\text{(non-TS)},\delta \eta_a) d(\delta \eta_a) =
\int \frac{\exp \{-\frac{1}{2} (y - M \delta \eta)^T C^{-1} (y - M \delta \eta)\}}{\sqrt{(2 \pi)^n |C|}} d(\delta \eta_a) \\
& = \frac{\exp \bigl\{-\frac{1}{2} y^T (C^{-1} - C^{-1} M (M^T C^{-1} M)^{-1} M^T C^{-1}) y \bigr\}}{\sqrt{(2 \pi)^{n-m} |C| |M^T C^{-1} M|}}
\equiv \frac{\exp \bigl\{-\frac{1}{2} y^T C' y \bigr\}}{\sqrt{(2 \pi)^{n-m} |C| |M^T C^{-1} M|}},
\end{aligned}
\label{eq:mmarglike}
\end{equation}
\end{widetext}
where $C = N + K^\text{(non-TS)}$.\footnote{To perform this integral, we remember the field-theoretical version of Gaussian integrals, $\int \mathrm{e}^{-\frac{1}{2} x^T A x + J^T x} \, dx = \sqrt{(2 \pi)^{\text{(size $A$)}} |A^{-1}|} \, \mathrm{e}^{\frac{1}{2} J^T A^{-1} J}$, and identify $A = M^T C^{-1} M$ and $J^T = y^T C^{-1} M$.}

A derivation of the van Haasteren--Levin result can also be given that remains closer in spirit to the logic of Gaussian processes. For that, we remember that the $\delta \eta_a$ can be seen as the weights of the basis functions $\phi^{(\mathrm{TS})}(t)$ (the columns of the design matrix $M$). We can then use Eq.\ \eqref{eq:loggplike} with $K^\mathrm{(TS)} = M \Sigma^\mathrm{(TS)} M^T$ and $\Sigma^\mathrm{(TS)} = \lambda I_p$, and take the limit $\lambda \rightarrow \infty$ corresponding to an infinitely vague prior for the TS weights:
\begin{widetext}
\begin{equation}
\begin{aligned}
\lim_{\lambda \rightarrow \infty} \log p(y|\theta^\text{(non-TS)}) &= 
\lim_{\lambda \rightarrow \infty} \biggl\{
-\frac{1}{2} y^T (C + M \lambda M^T)^{-1} y
-\frac{1}{2} \log |C + M \lambda M^T|
-\frac{n}{2} \log 2 \pi
\biggr\} \\
& = \lim_{\lambda \rightarrow \infty} \biggl\{
-\frac{1}{2} y^T C^{-1} y
+\frac{1}{2} y^T C^{-1} M (\lambda^{-1}I_p + M^T C^{-1} M)^{-1} M^T C^{-1} y \\
& \phantom{= \lim_{\lambda \rightarrow \infty} \biggl\{} -\frac{1}{2} \log |C| - \frac{1}{2} \log |\lambda^{-1}I_p + M^T C^{-1} M| - \frac{p}{2} \log \lambda -\frac{n}{2} \log 2 \pi
\biggr\} \\
& = -\frac{1}{2} y^T C' y -\frac{1}{2} \log |C| - \frac{1}{2} \log |M^T C^{-1} M|
-\frac{n - m}{2} \log 2 \pi \; (+ \; \text{infinite constant})
\end{aligned}
\label{eq:logmmarglike}
\end{equation}
In the second row of Eq.\ \eqref{eq:logmmarglike} we used the Woodbury formula and the matrix determinant lemma \cite{doi:10.1137/1031049}:
\begin{equation}
\label{eq:woodbury}
\begin{aligned}
(A + U W V^T)^{-1} &= A^{-1} - A^{-1} U (W^{-1} + V^T A^{-1} U)^{-1} V^T A^{-1}, \\
\det (A + U W V^T) &= \det (W^{-1} + V^T A^{-1} U) \det W \det A;
\end{aligned}
\end{equation}
\end{widetext}
we will have occasion to use these formulas repeatedly in the rest of this paper, and we will discuss their computational significance in Sec.\ \ref{sec:lowrank}.

van Haasteren and Levin \cite{2013MNRAS.428.1147V} later derived an alternative form for the $\delta \eta_a$-marginalized likelihood, which exploits the singular-value decomposition (SVD) $M = U \Sigma V^*$ \cite{trefethen1997numerical}. If $M$ is an $n \times n$ matrix, then $U$ and $V$ are orthogonal matrices of sizes $n \times n$ and $p \times p$ respectively, while $\Sigma$ is an $n \times p$ diagonal matrix. If we partition $U$ as $[F\,G]$, where $F$ comprises the first $p$ columns, we see that $F$ spans $\mathrm{range}(M)$, while $G$ spans the subspace orthogonal to $\mathrm{range}(M)$. Heuristically, we may reason that the projection $F F^T y$ of the residuals involves components that can be reabsorbed by a change in the $\delta \eta_a$, so these components are in effect \emph{unobserved} from the Gaussian-process perspective; a likelihood can then be written directly for the $(n - p)$-dimensional observable data vector $G G^T y$ (or more precisely, for the coefficients of the $y$ over the partial orthonormal basis given by the $G$ columns):
\begin{equation}
p(y|\theta^{(\text{non-TS})}) =
\frac{\exp\bigl\{-\frac{1}{2} y^T G (G^T C G)^{-1} G^T y\bigr\}}{\sqrt{(2\pi)^{n - p} |G^T C G|}}.
\label{eq:gmarglike}
\end{equation}
In App.\ \ref{sec:mgeq} we demonstrate that Eq.\ \eqref{eq:mmarglike} and
\eqref{eq:gmarglike} are indeed equivalent up to a multiplicative constant that does not affect Bayesian calculations.
%  MV: move this statement to appendix. ``with the only difference that we have marginalised over $\delta \eta^{\prime}$ instead of $\delta \eta$, with $M\eta=F \delta \eta^{\prime}$. The Jacobian of this transformation, $J = \bigl|(M^{T}M)^{-1}M^{T}F\bigr|$ is ignored in van Haasteren and Levin. This is acceptable, as this Jacobian is merely an uninformative multiplicative constant that does not change any details of the Bayesian inference.''

The computation of the ``$M$-matrix'' marginal likelihood [Eq.\ \eqref{eq:mmarglike}] is again dominated by the $O(n^3)$ inversion and determinant of the non-TS covariance $C$; it involves also $O(pn^2)$ and $O(p^2n)$ matrix--matrix multiplications, and the $O(p^3)$ inversion and determinant of $M^T C^{-1} M$, as well as negligible quadratic-order matrix--vector multiplications.
The computation of the ``$G$-matrix'' marginal likelihood [Eq.\ \eqref{eq:gmarglike}] is dominated by the $O\bigl((n-p)^3\bigr)$ inversion and determinant of the projected covariance $G^T C G$, and by $O\bigl((n - p)n^2\bigr)$ matrix multiplications (some of these can be avoided by storing the matrices $G^T K^{(A)}(\theta^{(A)}) G$ for varying values of $\theta^{(A)}$, and interpolating \cite{2013MNRAS.428.1147V}); it requires also the $O(np^2)$ SVD decomposition of $M$, which can be performed once and for all when we set up Bayesian inference.

\section{Low-rank formulations for correlated noise}
\label{sec:lowrank}

As we have seen so far, the bottleneck in the evaluation of Gaussian-process marginal likelihoods is the $O(n^3)$ computation of the inverse and determinant of the total covariance matrix $N + \sum_{(A)} K^{(A)}$. It is possible to improve on this situation by exploiting the specific structure of the individual covariance matrices.
For instance, the measurement-noise covariance matrix $N$ and certain of the $K^{(A)}$ are diagonal, with trivial $O(n)$ inverses. By contrast, other $K^{(A)}$ represent \emph{correlated noise}, and therefore \emph{a small number of effective degrees of freedom}; these matrices are usually \emph{severely rank deficient} (at least numerically, which is why they are so hard to invert), and they can be represented accurately by a \emph{truncated eigenvector expansion} $U S U^T$, where $U$ is $n \times l$ with $l \ll n$ \cite{golub2012matrix}.

Thus we are left with the task of computing the inverse of the sum of a diagonal matrix $D$ with a low-rank matrix $U S U^T$. This is where the Woodbury lemma \eqref{eq:woodbury} comes to the rescue. Indeed, its principal application in the literature is the \emph{low-rank update of an inverse}, which is just what we need:
\begin{multline}
    (D + U S U^T)^{-1} = \\ D^{-1} - D^{-1} U^{-1} (S^{-1} + U^T D^{-1} U)^{-1} U^T D^{-1}.
\label{eq:genwoodbury}
\end{multline}
We see that the matrix inversions in this reworked expression are those of $D$ (an $O(n)$ operation), $S$ (an $O(l^3)$ operation), and $(S^{-1} + U^T D U)$ (again $O(l^3)$), gaining us an impressive speedup. The corresponding lemma for the determinant is $|D + U S U^T| = |D| \, |S| \, |S^{-1} + U^T D^{-1} U|$, which reduces the original $O(n^3)$ computation to $O(n)$ and $O(l^2)$ operations.

The correlated-noise expansion is compatible with the $M$-matrix formulation of Sec.\ \ref{sec:timingmodel}, although the $O(p^3)$ inversion of $M^T C^{-1} M$ is still necessary (in addition, computing $M^T C^{-1} M$ itself is $O(pn^2)$. An alternative way to include the $M$-matrix marginalization is to replace $U$ in Eq.\ \eqref{eq:genwoodbury} with the concatenation $U^\prime=[M U]$, adopting an infinitely vague prior for the timing-model parameters, as we did in Eq.\ \eqref{eq:logmmarglike}.

With a little more work, the correlated-noise expansion is also compatible with the $G$-matrix formulation, where it leads to
\begin{multline}
y^T G \bigl(G^{T} (D + U S U^T) G\bigr)^{-1} G^T y = \\
y^T W y - y^T W U (S^{-1} + U^T W U)^{-1} U^T W y,
% \quad \text{with} \quad W = G (G^T D G)^{-1} G^T.
\label{eq:gmatrixwoodbury}
\end{multline}
with
\begin{equation}
W = G (G^T D G)^{-1} G^T.
\end{equation}

Now, the computation of the ``weight'' matrix $W$ involves an $O(n^3)$ inverse, which can be computed once and for all at the beginning of inference if its only dependence on the $\theta$ is a multiplicative constant such as an EFAC.
If the dependence of $W$ is more complicated, we can still avoid the $O(n^3)$ scaling by rewriting
\begin{equation}
G (G^T D G)^{-1} G^T = D^{-1} - D^{-1} F (F^T D^{-1} F)^{-1} F^T D^{-1},
\label{eq:quickw}
\end{equation}
(see Eq.\ \eqref{eq:laststep} in App.\ \ref{sec:mgeq}) where $F$ is the $n \times p$ orthogonal complement of $G$ (see Sec.\ \ref{sec:timingmodel}), and the required matrix inversions are therefore $O(n)$ and $O(p^3)$.

A computationally efficient expression for $W$ is also available when $D$ is the sum $a A + b B$ of two constant components, each multiplied by its own multiplicative hyperparameter (as in the case of single-backend/receiver EFAC and EQUAD noise). We can then diagonalize the two simultaneously with a non-orthogonal basis transformation:
\begin{equation}
    G^{T} D G = a G^{T} A G + b G^{T} B G = LV\bigl(a I + b Q\bigl) V^{T} L^{T},
%    \quad \text{with} \quad LL^{T} = G^T A G \quad \text{and} \quad VQV^{T} = L^{-1} G^{T} B G {L^{T}}^{-1},
\label{eq:twocomponentnoise}
\end{equation}
with
\begin{equation}
LL^{T} = G^T A G \quad \text{and} \quad VQV^{T} = L^{-1} G^{T} B G {L^{T}}^{-1},
\end{equation}
where $I$ is the identity matrix, $L$ is a lower-diagonal Cholesky decomposition \cite{trefethen1997numerical}, and $V Q V^{T}$ is an eigendecomposition, with $Q$ a diagonal matrix. The quantities required in Eq.\ \eqref{eq:gmatrixwoodbury} are now trivial to calculate:
\begin{multline}
    y G \bigl(G^{T} D G\bigl)^{-1} G^{T} y = \\ y^{T} G \left(V^{T} L^{T}\right)^{-1} \bigl(a I + b
    Q \bigl)^{-1} \left(L V\right)^{-1} G^{T} y,
\end{multline}
\begin{equation}
    \det \bigl(G^{T} D G\bigl) = \det \bigl(G^{T} A G\bigl) \det \bigl(a I + b Q \bigl).
\label{eq:twocomponentnoisetwo}
\end{equation}
Because we need to calculate $L V G^{T} y$ (or any other combination like $L V G^{T} U$) only once, the computational burden of the inverse is $O(n)$, and
evaluating Eq.\ \eqref{eq:gmatrixwoodbury} is $O(nl)$ and $O(l^3)$.

We note that, besides the low-rank expansions we outline in this section, another similar computational trick has been explored in \cite{2013MNRAS.429...55V}. Instead of using the Woodbury lemma to expand the low-rank representation of the covariance matrix, the data was compressed to a similar low-rank basis. The low-rank basis was not based on a frequency representation of the signal as we do in the next few sections, but on a high-fidelity basis derived from a Fisher-information matrix approximation of the likelihood. Linear interpolation of the compressed covariance matrices was subsequently used to obtain the covariance function for various model parameters. In Appendix \ref{sec:datacompression} we discuss linear data compression in the context of the more versatile frequency representation of signals, but in the rest of the paper we focus on the uncompressed data.

In the rest of this section we discuss the applications of low-rank expansions: in Sec.\ \ref{sec:fourier} for correlated timing noise; in Sec.\ \ref{sec:lowrankjitter}, for jitter-like noise in multifrequency datasets; in Sec.\ \ref{sec:epochaveraged}, to define a notion of epoch-averaged residuals. An alternative approach t, not based on low-rank expansions, but on linear data compression has also been used in the literature

\subsection{Low-rank expansions by Fourier sums}
\label{sec:fourier}

The Fourier-sum approach discussed in Secs.\ \ref{subsec:rednoise}--\ref{subsec:commonmode} for correlated timing noise, DM variations, and GWs leads directly to a low-rank approximation for the covariance matrix, which is obtained by setting, in the language of Eqs.\ \eqref{eq:fouriersum} and \eqref{eq:genwoodbury}, $U_{i\mu} = \phi^{(\mathrm{FM})}(t_i)$ and $D_{\mu \nu} = \Sigma^{(\mathrm{FM})}_{\mu\nu}$. As a reminder, $\mu$ ranges from 1 to $2q$, and indexes the Fourier basis functions $\cos(2 \pi f_\mu t_i)$ and $\sin(2 \pi f_\mu t_i)$ with $f_\mu$ a multiple of $1/T$, the inverse duration of the observation; the matrix $\Sigma^{(\mathrm{FM})}_{\mu\nu}$ is diagonal, with equal elements for each set of two bases of the same frequency. In the case of DM variations, the basis functions would be $(\nu_i/GHz)^{-2} \cos(2 \pi f_\mu t_i)$ and $(\nu_i/GHz)^{-2} \sin(2 \pi f_\mu t_i)$, following Eq.\ \eqref{eq:dmcovariance}.

If we are modeling correlated noise, DM variations, and GWs all together by way of low-rank expansions, we need to include a separate set of basis functions (and diagonal priors) for each. The resulting global $F$ matrix is obtained by stacking the individual $F$'s horizontally, and the global $\Sigma$ is the block-diagonal matrix of the individual $\Sigma$'s. However, since the bases for correlated noise and GWs are the same, except possibly for a different choice of $q$, the corresponding $F$ matrix needs to be included only once, and the two diagonal prior matrices can be summed. This means that the correlated-noise and GW hyperparameters will be correlated (partially or entirely, depending on the structure of the priors).

In the case of multi-pulsar analysis, the Fourier-sum modeling of GWs poses a challenge to the derivation of low-rank expressions. Each pulsar gets its own Fourier basis, but each such basis represents a Gaussian process that is correlated with the GW processes of the other pulsar. Let us label the residuals as $y_{ai}$, where $a$ indexes the pulsar and $i$ ranges over the residuals of each (which can be different numbers). If we use the same set $\{f_\mu\}$ of Fourier frequencies for all pulsars (based, e.g., on the duration of the longest dataset), the resulting multipulsar GW covariance is given by
\begin{equation}
\begin{aligned}
\label{eq:multipulsarcov}
K^{(\mathrm{GW})}_{ai\,bj} &= \sum_{a b \mu \nu} \Phi_{a \mu}(t_{ai}) 
\Sigma^{(\mathrm{GW})}_{a \mu\, b\nu}
\Phi_{b \nu}(t_{bj}) \\ &= \sum_{a b \mu \nu} \Phi_{a \mu}(t_{ai}) 
(\Sigma^{(\mathrm{GW})}_{\mu \nu} \, \zeta_{ab})
\Phi_{b \nu}(t_{bj})
\end{aligned}
\end{equation}
[see Eq.\ \eqref{eq:multicov}]. Now, $\Sigma^{(\mathrm{GW})}_{\mu \nu}$ is diagonal, but $\zeta_{ab}$ is dense, so its inverse is potentially expensive. If we order the Fourier coefficients in blocks corresponding to the $N$ pulsars, the matrix $\Sigma_{a \mu\, b\nu}$ appears to be made up of $N \times N$ blocks, each of which is a diagonal matrix. By contrast, if we order the coefficients in blocks corresponding to each $f_\mu$, then the matrix $\Sigma_{a \mu\, b\nu}$ is block-diagonal, with each block a dense matrix given by $\rho_\mu \gamma_{ab}$; thus, its inverse is just $\rho^{-1}_\mu \gamma^{-1}_{ab}$, which incurs an acceptable computational cost $O(q N^3)$.

In principle $\gamma^{-1}_{ab}$ could be saved and reused; however, if we are also modeling correlated noise with Fourier sums that share the same basis functions as the GWs, the resulting prior would be 
\begin{equation}
\label{eq:multipulsarprior}
\Sigma_{a \mu\, b\nu} = \Sigma^{(\mathrm{GW})}_{\mu \nu} \, \zeta_{ab} + \Sigma^{(\mathrm{red})}_{a, \mu \nu} \delta_{ab};
\end{equation}
in this case each block is given by $\rho^{(\mathrm{GW})}_\mu \gamma_{ab} + \rho^{(\mathrm{red})}_{a,\mu} \delta_{ab}$, and it must be inverted for each choice of $\rho^{(\mathrm{red})}_{a,\mu}$.

\subsection{Low-rank expansions for jitter-like noise}
\label{sec:lowrankjitter}

The covariance matrix corresponding to jitter-like noise, as described in Sec.\ \ref{subsec:equadjitter}, can be expressed \emph{exactly} as the low-rank expression
\begin{equation}
C_\mathrm{J} = U E U^T
\label{eq:jitterlowrank}
\end{equation}
where $E$ is a diagonal matrix with entries $J_e^2$ corresponding to squared amplitude of jitter-like noise at each epoch (usually the same for all epochs corresponding to measurements with the same receiver/backend), and where $U_{ie} = 1$ if measurement $i$ belongs to epoch $e$, 0 otherwise. If the residuals are sorted by epoch, the structure of the expansion is graphically obvious:
\begin{widetext}
\begin{equation}
\label{eq:exploder}
\small
\left(\begin{array}{cccc}
1 & & & \\
1 & & & \\
\vdots & & & \\
1 & & \\
& 1 & & \\
& 1 & & \\
& \vdots & & \\
& 1 & & \\
& & \ddots & \\
& & & 1 \\
& & & 1 \\
& & & \vdots \\
& & & 1
\end{array}\right)
\left(\begin{array}{cccc}
J_1^2 &       &        & \\
      & J_2^2 &        & \\
      &       & \ddots & \\
      &       &        & J_{n_e}^2
\end{array}\right)
\left(\begin{array}{ccccccccccccc}
1 & 1 & \cdots & 1 &   &   &        &   &        &   &   &        &   \\
  &   &        &   & 1 & 1 & \cdots & 1 &        &   &   &        &   \\
  &   &        &   &   &   &        &   & \ddots &   &   &        &   \\
  &   &        &   &   &   &        &   &        & 1 & 1 & \cdots & 1
\end{array}\right)
\end{equation}
\end{widetext}
%
% [MV: Bah, maybe this is too elementary to show. \rvh{yes, probably is...} MV: I'll keep it for the moment since I refer to it later.]
This representation can used together with the Fourier sums of Sec.\ \ref{sec:fourier} (for correlated-noise, DM-variation, and GWs) by stacking the $F$ and $U$ matrices as well as the priors. Otherwise, jitter-like noise can be kept in the matrix $D$ of Eqs.\ \eqref{eq:genwoodbury} and \eqref{eq:gmatrixwoodbury}. $D$ is then block diagonal (for sorted residuals) rather than diagonal, but its inverse can be computed very efficiently. Each block $D_e$ has the form of Eq.\ \eqref{eq:measnoise}:
\begin{equation}
D_e^{-1} = (N_e + J_e^2 u_e u_e^T)^{-1},
\label{eq:jitterinverseone}
\end{equation}
where $(N_e)_{ij} = (E_e^2 n_i + Q_e^2) \delta_{ij}$, with indices ranging over the epoch only, and $u_e^T = (1,1,\ldots,1)^T$. By Woodbury's lemma,
\begin{equation}
D_e^{-1} = N_e^{-1} - \frac{N_e^{-1} u_e u_e^T N_e^{-1}}{\alpha_e}, \;
\text{with} \; \alpha_e = J_e^{-2} + u_e^T N_e^{-1} u_e,
\label{eq:jitterinversetwo}
\end{equation}
which is $O(b^2)$, with $b$ the dimension of the block. Altogether $D^{-1}$ can be computed in $O(n\bar{b})$, with $\bar{b}$ the average number of residuals in an epoch.

\subsection{Low-rank expansion of epoch-averaged residuals} 
\label{sec:epochaveraged}

A low-rank expansion can also be used to define a statistically principled notion of \emph{epoch-averaged residual} for multifrequency datasets such as NANOGrav's [Michele, unpublished NANOGrav memo]. The idea is to write the total Gaussian-process covariance matrix as $N + C = N + U \tilde{C} U^T$, where the $n \times n$ matrix $N$ includes the measurement noise components (such as EFAC and EQUAD noise) that are independent for each residual, while the $n_e \times n_e$ matrix $\tilde{C}$ (with $n_e$ the number of epochs) describes components such as jitter-like noise, correlated noise, and GWs\footnote{DM fluctuations require a slightly more complicated description where $U$ gains $n$ rows, with the same structure as its those of Eq.\ \eqref{eq:exploder}, but each multiplied by $\nu^{-2}_i$.} that depend only on the observation time of each epoch, and are therefore entirely correlated among residuals in the same epoch; thus, the ``exploder'' matrix $U$ has the same structure as in Eq.\ \eqref{eq:exploder}.

A Woodbury expansion yields the likelihood in the form
\begin{widetext}
\begin{equation}
\begin{aligned}
& -\frac{1}{2} y^T (N + U \tilde{C} U^T)^{-1} y - \frac{1}{2} \log |N + U \tilde{C} U^T| \\
=&
-\frac{1}{2} y^T N^{-1} y - \frac{1}{2} \log |N| 
+ \frac{1}{2} y^T N^{-1} U (\tilde{C}^{-1} + U^T N^{-1} U)^{-1} U^T N^{-1} y
- \frac{1}{2} \log |\tilde{C}| |\tilde{C}^{-1} + U^T N^{-1} U| \\
=& -\frac{1}{2} \tilde{\chi}^2  - \frac{1}{2} \log |N|
+ \frac{1}{2} \tilde{y}^T (\tilde{C}^{-1} + X)^{-1} \tilde{y}
- \frac{1}{2} \log |\tilde{C}| |\tilde{C}^{-1} + X|,
\end{aligned}
\label{eq:epochaveraged}
\end{equation}
\end{widetext}
where we have neglected logarithms of $2\pi$. In Eq.\ \eqref{eq:epochaveraged} the $n$-dimensional vector of residuals $y$ is replaced by the $n_e$-dimensional vector of epoch-averaged residuals $\tilde{y} = U^T N^{-1} y$. Thus, in principle a full multifrequency dataset can be condensed into $\tilde{y}$, plus the white-noise $\tilde{\chi}^2$ of the observation and the $n_e \times n_e$ matrix $X = U^T N^{-1} U$ of averaged measurement noise. The marginalization over the timing-model parameters can also by accommodated, in the $G$-matrix formulation of Eq.\ \eqref{eq:gmarglike}, by redefining $\tilde{y} = U^T W y$, $\tilde{\chi}^2 = \tilde{y}^T W \tilde{y}$, and $X = U^T W U$, with $W = G (G^T D G)^{-1} G^T$.

Unfortunately epoch averaging is not useful in practice because the measurement-noise matrix $N$ is usually a function of several hyperparameters (the EFACs and EQUADs for the various backend/receiver combinations), so the full set of residuals must be carried along throughout the analysis to recompute $\tilde{y}$, $\tilde{\chi}^2$, and $X$ as the hyperparameters change. If a single EFAC and EQUAD describe the entire dataset, then epoch-averaged residuals can be used by way of the two-component expansion of Eq.\ \eqref{eq:twocomponentnoise}.

% [MV: although... in a Gibbs scheme all the steps except the ``white-noise'' step could deal with epoch-averaged residuals, and the white-noise step could recompute $\tilde{y}$, $\tilde{\chi}^2$, and $X$. Would this save anything. \rvh{It depends. For a single pulsar, sure. For a full array, with a GWB search, I'd rather not include the quantization matrix in the red-noise Woodbury epansion. That greatly increases dimensionality of low-rank approx. That's why I want to either include it with a separate Gibbs block, using epoch-averaged jitter-only residuals as quadratic parameters, or (better) include it with high-performance operations in the white-noise step.}]

\section{Quasi-Gibbs schemes for Bayesian inferences on pulsar-timing datasets} \label{sec:sampling}

	Performing Bayesian inference for model parameters and hyperparameters requires the exploration of a high-dimensional parameter space to build a representation of the posterior parameter distributions.
	Reducing the number of search parameters by marginalizing over some of them analytically, as we discussed in Sec.\ \ref{sec:timingmodel}, can be part of the solution, but it is not the entire story.
	The reason is that stochastic methods such as Markov Chain Monte Carlo (MCMC) are typically used to explore the space of the remaining parameters, so the efficiency of an inference scheme depends crucially on the number of likelihood evaluations required to sample the posteriors broadly and accurately enough, as well as the computational cost of an individual likelihood evaluations.
	The need to choose wisely is especially pointed for large datasets such as the upcoming IPTA data releases, which may contain many tens of thousands TOAs, requiring (in principle) the inversion of matrices with billions of elements.
	
	MCMC methods explore parameter posteriors by using (in effect) a guided random walk: they generate a sequence of \emph{samples} whose distribution converges asymptotically to the posterior. The rate of convergence, regardless of the dimension of parameter space, can be characterized as $1/\sqrt{N}$, where $N$ is the number of samples (strictly speaking, it is the fractional error of integrated quantities such as $\int \phi(x) p(x) \mathrm{d}x$	that scales as $\langle \phi \rangle / \sqrt{N}$, with $\langle \phi \rangle$ is the variance of the function $\phi(x)$). The $N$ in this scaling, however, is really the number of \emph{statistically independent} samples, which is related to the length of the chain by a multiplicative constant that depends on the dimension of parameter space, on the structure of the posterior, and on the particular scheme used to generate the chain. For an actual chain, the multiplicative constant is characterized well by the \emph{sample autocorrelation function} (ACF), defined as
	\begin{equation}
    	\mathrm{ACF}_{t}(x)\frac{\frac{1}{N-1}\sum_{i=1}^{N-t}\left(x_i-\bar{x}\right)\left(x_{i+t}-\bar{x}\right)}
                {\frac{1}{N-1}\sum_{i=1}^{N}\left(x_i-\bar{x}\right)^{2}},
        \label{eq:acf}
    \end{equation}
	where $x$ and $N$ are the vector and number of samples, $\bar{\cdot}$ indicates the sample mean, and $t$ is the sample lag. The lag at which the ACF drops by a factor $e$ is known as the exponential autocorrelation length \cite{liu2001monte} ; sampling schemes that yield lower autocorrelation lengths for all parameters require correspondingly fewer samples to achieve the same accuracy.
	
	In this section we present two schemes, inspired by Gibbs sampling (see Sec.\ \ref{sec:gibbs} below), that result in much lower autocorrelation lengths than the state-of-the-art methods currently in use. In Sec.\ \ref{sec:gibbsspectrum} we introduce a scheme optimized for spectral estimation (i.e., for correlated-noise and GW models with free Fourier-sum coefficients); in Sec.\ \ref{sec:modifiedgibbs} we describe a modified scheme that is useful for model spectra (e.g., power-law correlated noise and GWs).

    \subsection{Gibbs sampling for pulsar-timing analysis} \label{sec:gibbs}
    The simplest MCMC schemes are based on the \emph{Metropolis--Hastings rule}: each new sample in the sequence $\{\theta^{(n)}\}$ is generated by first \emph{proposing} a new parameter vector $\theta^{(n+1)}$ from a proposal distribution $q(\theta^{(n+1)}|\theta^{(n)})$ (which often describes a local perturbation), then \emph{accepting} it with probability given by the Metropolis--Hastings ratio
    \begin{equation}
    \frac{p(\theta^{(n+1)}|\mathrm{data})}{p(\theta^{(n)}|\mathrm{data})} \times
    \frac{q(\theta^{(n)}|\theta^{(n+1)})}{q(\theta^{(n+1)}|\theta^{(n)})}.
    \end{equation}
    The resulting \emph{detailed balance} (the fact that the flow of samples between two locations in parameter space is proportional to the ratio of the posteriors) guarantees the existence of an equilibrium distribution. If the proposal is such that the chain is \emph{ergodic} (it can reach any corner of parameter space), convergence is assured in the limit. Choosing the proposal distribution smartly \cite[see, e.g.][]{Roberts1997} is paramount to achieving good chain \emph{mixing} (low autocorrelation lengths).
   		
		In a \emph{Gibbs scheme} \cite{10.1109/TPAMI.1984.4767596}, by contrast, at each step one modifies only a subset of parameters (often just one), and does so by drawing the new value directly from the \emph{conditional} probability distribution of the modified parameters given the unmodified ones. If the blocks of parameters that are modified together are chosen to minimize correlations between blocks, the resulting chain mixing is very good, because all parameters are in effect drawn from the global posterior, except for the effects of residual inter-block correlations.
		
		For pulsar-timing analysis, the opportunity of using Gibbs sampling is motivated by a crucial observation on the full unmarginalized likelihood for the case of Fourier-sum correlated noise and GWs (see Sec.\ \ref{sec:fourier}):
        \begin{equation}
        \begin{aligned}
            p(y|\theta) = & \,
            \frac{\exp \{-\frac{1}{2} (y - M \delta \eta - Fa)^T N^{-1} (y - M
            \delta \eta - Fa)\}}{\sqrt{(2 \pi)^n |N|}} \\ & \times
            \frac{\exp \{-\frac{1}{2} a \Sigma^{-1} a\}}{\sqrt{(2 \pi)^p |\Sigma|}}.
            \label{eq:fulllik}
        \end{aligned}
        \end{equation}
		[here the $a$ and $F$ are the weights and basis matrix of the Fourier-sum Gaussian processes, and $\Sigma$ encodes their priors; the $\delta \eta$ and $M$ are the timing-model parameter errors and design matrix, and the infinitely vague prior is implicit.]
		Equation \eqref{eq:fulllik} can be interpreted as a conditional probability for the $a$ and $\delta \eta$ given the hyperparameters that define $N$ and $\Sigma$. Furthermore, the conditional probability is Gaussian, which makes it easy to sample from it, as wee see below. If the hyperparameters that define $\Sigma$ are given in the ``spectral--estimation'' form $\Sigma_{\mu \nu} = \rho_\mu \delta_{\mu \nu}$ (see Eq.\ \eqref{eq:fouriersum}), then the $\rho_\mu$ can also be drawn easily from their conditional posteriors, given the weights and the other hyperparameters.
		Last, Eq.\ \eqref{eq:fulllik} requires the inversion of diagonal matrices only (an $O(n)$ operation), so it can be evaluated very efficiently.
		
		The reason why the full Fourier-sum likelihood has not been used so far in pulsar-timing analysis is that the resulting increase in computational efficiency is outweighed by the increased autocorrelation lengths in MCMC schemes that evolve the hyperparameters together with the weights in perturbative fashion. A quasi-Gibbs, blocked sampling scheme overcomes this problem (the scheme is not quite Gibbs because we still need Metropolis--Hastings updates for the hyperparameters that appear nontrivially in the likelihood). We describe it in the next section.        

	\subsection{Quasi-Gibbs, blocked sampling scheme for spectral estimation}
	\label{sec:gibbsspectrum}
	
		In this sampling scheme we successively modify the values of blocks of parameters, holding all the others fixed (hence the scheme is \emph{blocked}). The choice of blocks aims at two goals: the covariance between parameters in separate groups should be minimized to improve the ACF, and it should be possible to sample directly from the conditional probabilities for each block, or at least to evaluate them cheaply. Thus, we choose the following groups:
        \begin{enumerate}
            \item \emph{Quadratic parameters}, consisting of the timing-model parameter errors $\delta \eta$ and the Fourier coefficients $a$ for both correlated noise, GWs, and DM variations. (For a single pulsar, GWs would be degenerate with correlated noise.) 
            \item Hyperparameters describing white noise, and optionally jitter-like noise, using Eqs.\ \eqref{eq:jitterinverseone} and \eqref{eq:jitterinversetwo}; jitter-like noise could also be modeled with quadratic parameters, per Eq.\ \eqref{eq:jitterlowrank}.
            \item Hyperparameters describing priors for correlated-noise and GW Fourier coefficients.
            \item Hyperparameters describing priors for DM-variation Fourier coefficients.
        \end{enumerate}
        We cycle through these four steps, resampling the parameters in each block while holding the others fixed to their most recent value; at the end of each cycle we obtain a full Markov-chain sample. We now discuss each step in detail.

        \emph{1. Sampling the quadratic parameters.}
        % \label{sec:gibbsquadratics}
        As mentioned before, the quadratic parameters are the weights of the basis functions $\phi_\mu^{(\mathrm{TS})}(t_i)=M_{i\mu}$ and $\phi^{(\mathrm{FM})}_{\mu}(t_i)=F_{i\mu}$ of the timing model and the correlated noise respectively. We denote them collectively as $w^T = (\delta \eta^T, a^T)$ and with $\Phi = (M;F)$. Fixing all the hyperparameters $\theta$, the log-posterior probability of the $w$ can be rewritten as
        \begin{equation}
        \begin{aligned}
        	& \log P(w | y, \theta, \mathrm{GP}) = \\ &\quad-\frac{1}{2}\left(w - Q^{-1}
            \Phi N^{-1}y \right)^{T} 
            Q \left(w - Q^{-1}\Phi N^{-1}y \right) \\ &\quad- \frac{1}{2}\log\det Q
            + \mathrm{const}
            \label{eq:group1conditional}
        \end{aligned}
        \end{equation}
        with
        \begin{equation}
            Q = \Phi N^{-1} \Phi^{T} + \Sigma^{-1},
            \label{eq:Qaux}
        \end{equation}
        where $N$, $\Sigma$, and the additive constant are functions of the hyperparameters and of the residuals, and where we interpret $\Sigma^{-1}$ in the broad sense explained in Sec.\ \ref{sec:timingmodel}: by assuming an infinitely vague prior for the timing-model parameters, we set $\Sigma^{-1}$ to zero in their subspace. Equation \eqref{eq:group1conditional} states that the quadratic parameters $w$ are distributed according to a multivariate normal distribution with mean $\bar{w} = Q^{-1}\Phi N^{-1}y$ and covariance $Q^{-1}$. We can draw from this distribution by computing $w_\mathrm{new} = \bar{w} + L \epsilon$, with $\epsilon$ a vector of zero-mean, unit-norm, uncorrelated normal deviates (see, e.g., \cite{1992nrca.book.....P}), and $L$ a square root of $Q^{-1}$ (i.e., $L L^T = Q^{-1}$). For numerical stability, we first evaluate $Q^{-1}$ with a QR decomposition \cite{trefethen1997numerical}, then use an SVD decomposition \cite{trefethen1997numerical} to compute the square root.

		In multipulsar datasets, the effects of GWs on the timing residuals of different pulsars are correlated [see Eq.\ \eqref{eq:multipulsarcov}]; thus, so are the posterior distributions of the GW Fourier coefficients for each pulsar [by way of Eq.\ \eqref{eq:multipulsarprior}]. If we were to use the procedure that we have just outlined to draw new GW quadratic parameters, we would have to do so for all the pulsars at once, which can be very computationally expensive. Instead, the step can be performed separately for each pulsar $a$ by conditioning the corresponding $w_a$ on the most recent $w_b$ for all $b \neq a$. Expanding Eq.\ \eqref{eq:fulllik} for a prior matrix $\Sigma$ that includes cross terms between pulsars and collecting all the terms that involve the $w_a$ results in the conditional probability
        \begin{equation}
        \begin{aligned}
            &\log P(w_a | w_{b \neq a}, y, \theta, \mathrm{GP}) = \\ &\quad-\frac{1}{2}\left(w_a - Q_{a}^{-1} z_a \right)^{T} Q_a \left(w_a - Q_a^{-1} z_a \right)  \\ &\quad- \frac{1}{2}\log\det Q_a
                + \mathrm{const}(\theta,w_{b \neq a}),
            \label{eq:gwgroup1conditionalquad}
		\end{aligned}
        \end{equation}
        where
        \begin{equation}
        \begin{gathered}
                z_a = \Phi_a N_a^{-1} y_a + \sum_{b \neq a} (\Sigma^{-1})_{ab} w_b, \\
                Q_a = \Phi_a N_a^{-1} \Phi_a^{T} + (\Sigma^{-1})_{aa},
                \label{eq:gwgroup1aux}
        \end{gathered}
        \end{equation}
        from which $w_a$ can be drawn directly with the covariance-square-root procedure.
        Because of the structure of the multipulsar prior [Eq.\ \eqref{eq:multipulsarprior}], computing the submatrices $(\Sigma^{-1})_{aa}$ and $(\Sigma^{-1})_{ab}$ does not require the inversion of the full $\Sigma$, but only of each pulsar block, which is much cheaper.

        \emph{2. Sampling the white-noise hyperparameters.}
        % \label{sec:gibbswhitenoise}
        The conditional probability for the white-noise hyperparameters $\theta_w$ that determine $N$ in Eq.\ \eqref{eq:fulllik} is very simple:
        \begin{multline}
            \log P(\theta_w | y_\mathrm{red}, w, \theta_p, \mathrm{GP}) = \\ -\frac{1}{2} y_\mathrm{red}^T N^{-1} y_\mathrm{red} - \frac{1}{2}\log\det N - \log p(\theta_w),
            \label{eq:group3conditional}
        \end{multline}
        where the reduced residuals $y_\mathrm{red} - \Phi w$ are obtained by subtracting the most recent realization of the quadratic-parameter Gaussian processes from the residuals, and where $p(\theta_w)$ is the prior for the $\theta_w$. We cannot draw directly from this distribution, but we can approximate such a draw by performing a sequence of perturbative Metropolis--Hastings updates (in effect, a small MCMC run) for the $\theta_w$. Because of the form of Eq.\ \eqref{eq:group3conditional}, this is not costly.
                
        \emph{3 and 4. Sampling the Fourier-sum hyperparameters.}
        % \label{sec:gibbsspectrum}
        The following description applies to the Fourier-sum hyperparameters for correlated timing noise and GWs, and for DM variations. We denote either set as $\theta_p$. The conditional probability for the $\theta_p$, fixing everything else, is given by
        \begin{multline}
            \log P(\theta_p | y, \theta_w, w, \mathrm{GP}) = \\ -\frac{1}{2} w^T
            \Sigma^{-1} w - \frac{1}{2}  \log\det \Sigma - \log p(\theta_p),
            \label{eq:group2conditional}
        \end{multline}
        where, again, $\Sigma^{-1}$ is identically zero in the subspace of the timing model parameters, which do not appear in this equation. (This is not an inherent restriction of our scheme, but it is our choice.)
        
        The ``spectral-estimation'' model discussed in Sec.\ \ref{subsec:rednoise} includes an independent variance parameter $\rho_\mu$ on the diagonal of $\Sigma$ for each modeled frequency. Each $\rho_\mu$ applies to a cosine and and a sine mode; we will denote their weights as $a_\mu$ and $b_\mu$. If we adopt $1/\rho_\mu$ Jeffreys priors for each $\rho_\mu$ \cite{J03-PTLOS}, Eq.\ \eqref{eq:group2conditional} becomes fully separable, and we can write
        \begin{equation}
            P(\rho_\mu | a_\mu, b_\mu, \theta_w, \mathrm{GP}) = 
            \frac{\left(a_\mu^2+b_\mu^2\right)
            \exp\left( -\frac{1}{2}\frac{a_\mu^2+b_\mu^2}{\rho_\mu} \right)}
            {\rho_\mu^2}.
            \label{eq:group2individual}
        \end{equation}
        We can draw samples from this distribution analytically, even if we adopt a proper Jeffreys prior with compact support $\rho_{\mu,\mathrm{min}} < \rho_\mu < \rho_{\mu,\mathrm{max}}$. To do so, we pick $\eta$ uniformly in the interval
        $\left[0, 1-\exp\left(\tau/\rho_{\mu,\mathrm{max}} - \tau/\rho_{\mu,\mathrm{min}}\right)\right]$, with $\tau = (a_\mu^2+b_\mu^2)/2$, and we compute
        \begin{equation}
        	\rho_{\mu,\mathrm{new}} =
            \frac{\tau}{\tau/\rho_{\mu,\mathrm{max}} - \log\left(1-\eta\right)}.
            \label{eq:group2insam}
        \end{equation}
        With a more general prior $p(\theta_p)$, we can still use the small-MCMC strategy discussed above for $\theta_w$.

		This scheme is analog to augmented/missing-data methods used in machine learning \cite{liu2001monte}: if we think of the Fourier coefficients as \emph{unobserved data} rather than model parameters, then at the beginning of each cycle we are in effect \emph{imputing} their values (according to their conditional probability with the current hyperparameters) to ``complete'' the dataset, and evaluate model-parameter likelihoods with greater convenience.

		In actual use, this scheme turns out to be very efficient, with extremely low autocorrelation lengths (see Sec.\ \ref{sec:simspectrum}). This is because nearly all the parameters in the different blocks turn out to be nearly uncorrelated; the only significant correlations are between the quadratic-spindown timing-model parameter and the lowest Fourier coefficients, which do not increase the overall autocorrelation length significantly.
		In addition, the Fourier coefficients are also effectively uncorrelated among themselves, because the corresponding modes are approximately orthogonal (they would be exactly orthogonal if the TOAs were sampled regularly). This does not matter to their update step, since we are drawing from the joint posterior; however, this noncorrelation helps chain mixing, because it means that each pair of $(a_\mu,b_\mu)$ interacts (and correlates) with a single $\rho_\mu$ that is updated in a different block.
		
\begin{figure}
            \includegraphics[width=\columnwidth]{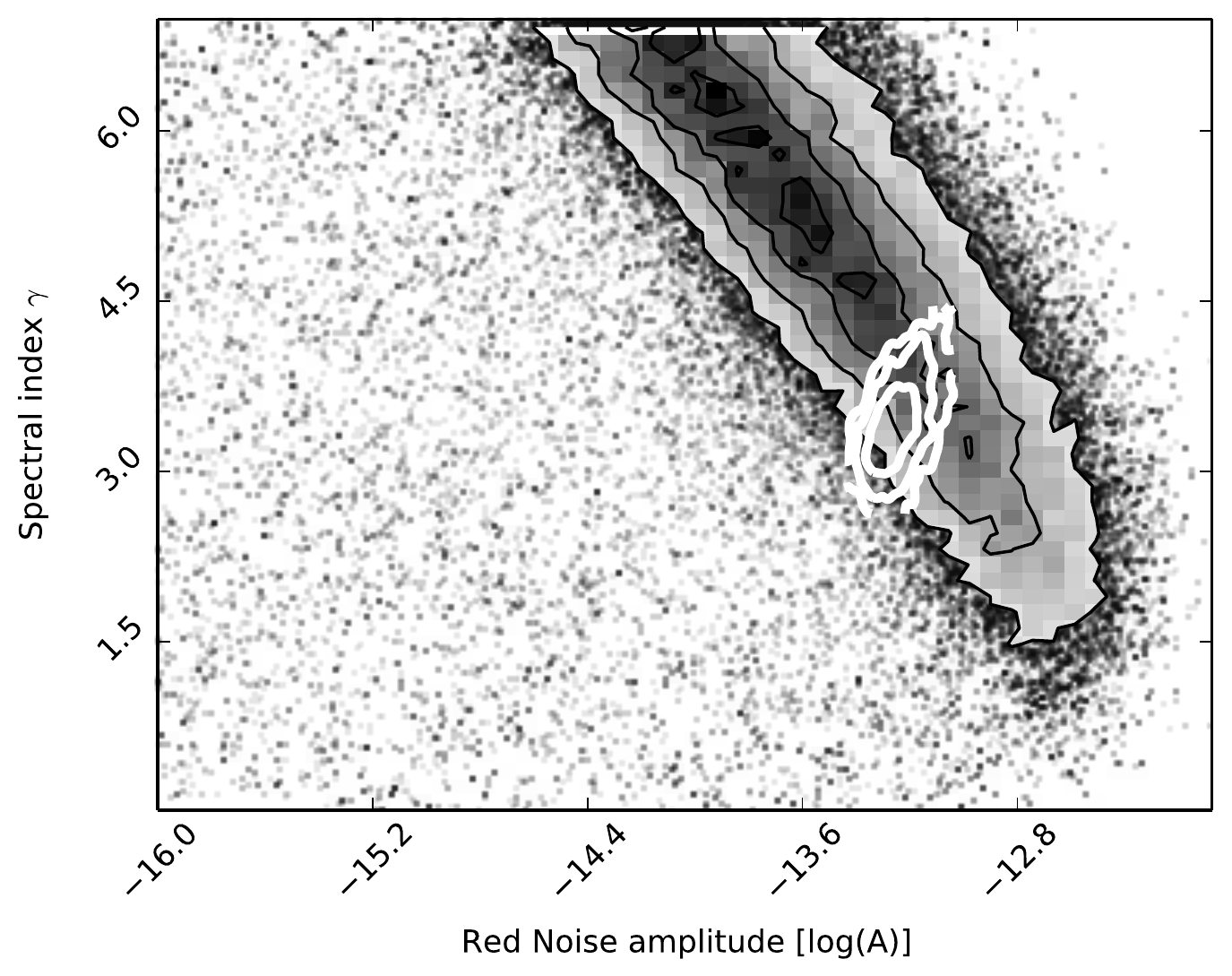}
            \caption{Comparison of the correlation profiles for the correlated-noise amplitude and spectral slope parameters in a full MCMC run (larger density profile) and as ``seen'' in a correlated-noise hyperparameter block update, where the Fourier coefficients are fixed to specific ``imputed'' values. The runs were performed on NANOGrav's 5-year J1910+1256 datasets, which includes DM corrections as part of the timing model \cite{2013ApJ...762...94D}.}
            \label{fig:J1910minichain}
        \end{figure}

		However, if we apply quasi-Gibbs scheme to a model of correlated noise where the Fourier and timing-model coefficients are correlated more strongly through the hyperparameters (as in model with power-law spectral densities), the autocorrelation lengths increase sharply. To illustrate this problem, in Fig.\ \ref{fig:J1910minichain} we show the correlation profile of the correlated-noise power-law parameters (amplitude and spectral slope), as estimated in a standard marginalized-poster MCMC, together with the much smaller conditional-correlation profiles (white curves) that is ``seen'' in one of the hyperparameter block updates of the quasi-Gibbs scheme, where all the Fourier coefficients are fixed to specific values. The limited extension of the effective correlation profiles greatly increases the autocorrelation times of the hyperparameters in the quasi-Gibbs chain.
		
		\subsection{Collapsed quasi-Gibbs sampling scheme for modeled spectra}
		\label{sec:modifiedgibbs}
		
		To improve this behavior, we need to sample the Fourier coefficients and their hyperparameters simultaneously. This is what we do in the modified scheme described here, which trades some computational efficiency for shorter autocorrelation lengths.
		This scheme has the same four steps as the quasi-Gibbs scheme of the last section, but we modify the step 3/4 where we update the hyperparameters of the modeled spectra. For these we adopt the following procedure: a) we first draw new hyperparameters $\theta_p$ from a perturbative proposal; b) we then generate new quadratic parameters $w_p$ directly from their conditional posterior given the new $\theta_p$; finally c) we accept the new $(\theta_p,w_p)$ according to the Metropolis--Hastings rule. 
		
		The Metropolis--Hastings ratio for the entire step is then
		\begin{equation}
		\frac{p(\theta_p^{(n+1)},w_p^{(n+1)}|y,\ldots)}{p(\theta_p^{(n)},w_p^{(n)}|y,\ldots)}
		\times
		\frac{q(w_p^{(n)}|w_p^{(n+1)}) q(\theta_p^{(n)}|\theta_p^{(n+1)})}{q(w_p^{(n+1)}|w_p^{(n)})q(\theta_p^{(n+1)}|\theta_p^{(n)})},
		\end{equation}
		where we do not indicate the dependence of the probabilities on all the hyperparameters and coefficients that are not updated in this step. However, since the proposal for $w_p$ is just its conditional given the new hyperparameters,
		\begin{equation}
		q(w_p^{(n+1)}|w_p^{(n)}) = p(w_p^{(n+1)}|\theta_p^{(n+1)},y,\ldots),
		\end{equation}
		and since the overall posterior probability can be factorized as
		\begin{equation}
		p(\theta_p,w_p|y,\ldots) = p(\theta_p|y,\ldots) p(w_p|\theta_p,y,\ldots),
		\end{equation}
		where $p(\theta_p|y,\ldots)$ is marginalized over the $w_p$, the Metropolis--Hastings ratio simplifies (\emph{collapses}) to
		\begin{equation}
		\frac{p(\theta_p^{(n+1)}|y,\ldots)}{p(\theta_p^{(n)}|y,\ldots)}
		\times
		\frac{q(\theta_p^{(n)}|\theta_p^{(n+1)})}{q(\theta_p^{(n+1)}|\theta_p^{(n)})},
		\end{equation}
		with
		\begin{widetext}
		\begin{equation}
		\log p(\theta_p|y,\ldots) = -\frac{1}{2} y^T \left(N^{-1}-N^{-1}\Phi\left(\Phi^{T}N^{-1}\Phi+\Sigma^{-1}\right)^{-1} \Phi^{T}N^{-1}\right)y - \frac{1}{2} \log | \Phi^{T}N^{-1}\Phi+\Sigma^{-1}|
		+ \mathrm{const}.
		\end{equation}
		\end{widetext}
		Thus we are just taking a Metropolis--Hastings step over the $\theta_p$ using the marginalized posterior, and we can wait to draw new $w^{(n+1)}_p$ from the conditional probability given the $\theta_p^{(n+1)}$ only if the step is accepted (conveniently, we already have the appropriate $Q^{-1}$ covariance to do so).
		
		In addition to reworking step 3/4, we need also to adjust the parameter blocks, by including among the quadratic parameters that are updated also the timing-model parameters that are significantly covariant with them (i.e., the quadratic spindown in the correlated-noise block, and the DM parameter\footnote{And when not accurately modeling the lowest DM variation frequencies, also the first and second time derivatives of the DM.} in the DM-variation block).
		From a computational-cost standpoint, this scheme is comparable to an MCMC based on the fully marginalized posterior. However the resulting autocorrelation lengths are much improved by the blocked updates of uncorrelated parameter subsets (see Sec.\ \ref{sec:simgwb}).

\section{Tests of the quasi-Gibbs schemes using mock data} \label{sec:simulations}
	In this section we test the performance (and basic correctness) of our quasi-Gibbs sampling schemes using simulated timing residuals, which we obtain using the \texttt{libstempo} interface \cite{libstempo} to the \texttt{Tempo2} timing package \cite{tempo2web}. 
	In Sec.\ \ref{sec:simspectrum} we compare the spectral-estimation quasi-Gibbs method of Sec.\ \ref{sec:gibbsspectrum} with a standard MCMC method, applying both to a single-pulsar dataset. In Sec.\ \ref{sec:simgwb} we compare the more general quasi-Gibbs method of Sec.\ \ref{sec:modifiedgibbs} with again a standard MCMC method, applying both a multipulsar dataset that contains a GW background.

    \subsection{Test of spectral-estimation quasi-Gibbs scheme} \label{sec:simspectrum}
\begin{figure}
        	\includegraphics[width=\columnwidth]{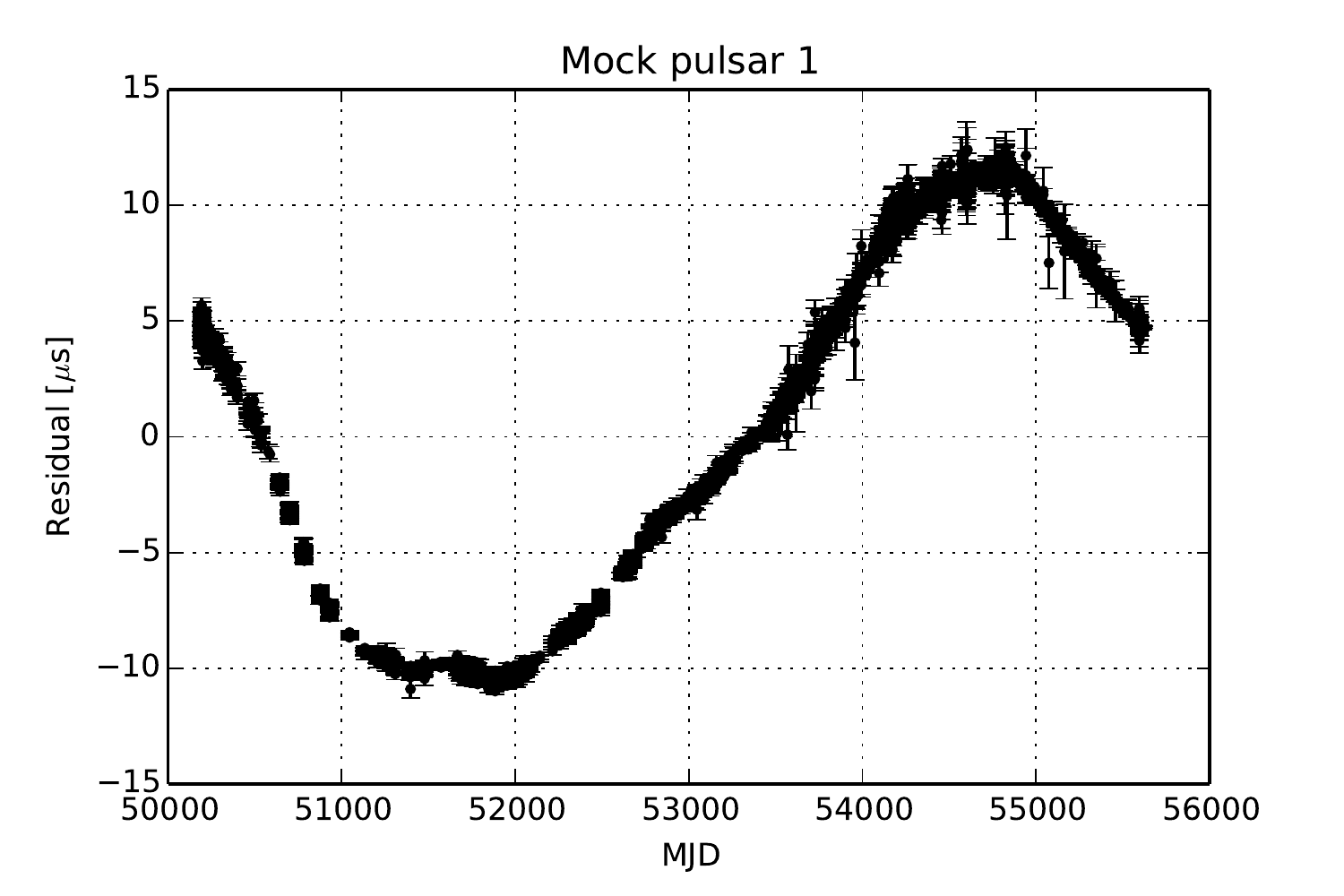}
            \includegraphics[width=\columnwidth]{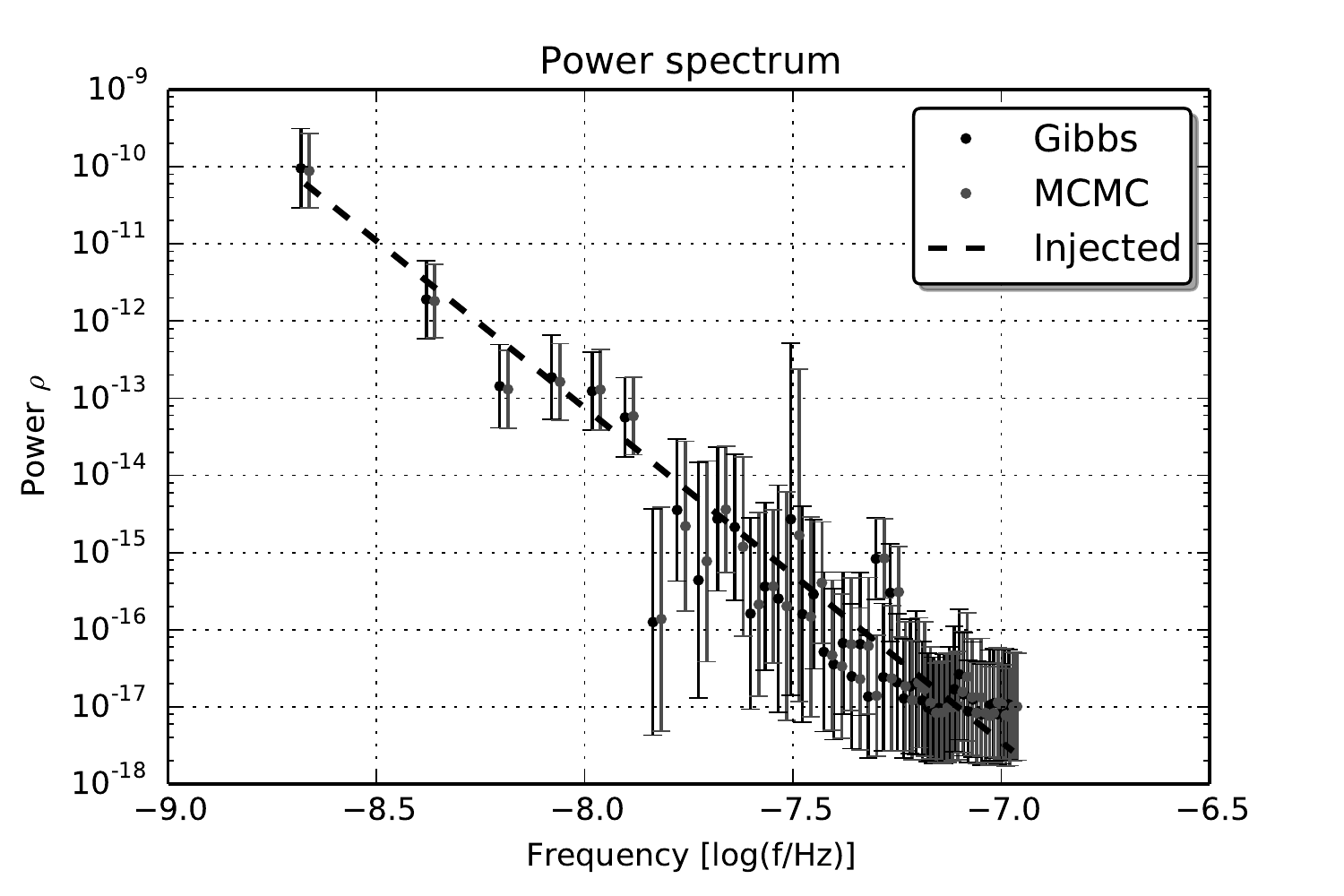}
            \caption{Mock data and spectral estimation in the test of the quasi-Gibbs scheme. Top: Mock residuals for pulsar J$0437$-$4715$ used in the test of Sec.\ \ref{sec:simspectrum}. We generated 1,500 TOAs over a time span of 6,000 days, injecting power-law timing noise [Eq.\ \eqref{eq:powerlaw}] with parameters $A=3\times10^{-14}$, and $\gamma=4.33$. Bottom: Recovered Fourier-mode variances for the Metropolis and quasi-Gibbs samplers. The error bars show 1-$\sigma$ standard deviations, and the spectrum of injected noise is shown as the dashed line. \label{fig:mockpsr1res}}
        \end{figure}

        The single-pulsar mock dataset for this test is based on the timing model of pulsar $J0437$-$4715$ in the ATNF pulsar catalog \cite{atnf}. This is one of the IPTA pulsars with the lowest TOA uncertainty, and it has been observed regularly. We generate timing residuals with real-world characteristics: the TOAs are sampled unevenly (in the MJD interval 50,000-56,000), they reflect strong timing noise, and their TOA uncertainties are varying.
		As typical for actual data collected with ever-evolving observation systems, we partition the dataset in 15 blocks, all corresponding to different hardware, each with a different EFAC and EQUAD. The residuals for this mock dataset are shown in Fig.\ \ref{fig:mockpsr1res}.
        
        In our test we determine the power spectral density of injected noise, in the style of Eq.\ \eqref{eq:sumovermodes} and Ref.\ \cite{2013PhRvD..87j4021L}, using two sampling schemes: a ``vanilla'' adaptive Metropolis MCMC sampling method (as described in the appendix of \cite{2014arXiv1404.1267A}), and the quasi-Gibbs method of Sec.\ \ref{sec:gibbsspectrum}.
        With both methods we adopt the same noise model: white noise with $15 + 15$ EFAC and EQUAD hyperparameters, plus correlated noise described by 50 Fourier modes at frequency multiples of $1/T$, with 50 independent variance parameters describing the spectral density. For the adaptive Metropolis sampler, the posterior is marginalized analytically over all quadratic parameters, so the total dimension of parameter space is 80.  For the quasi-Gibbs sampler, the unmarginalized posterior is a function of $214$ parameters: $30$ white-noise hyperparameters, $50$ spectral-density prior variances, $100$ Fourier coefficients, and $34$ timing model parameters, of which $12$ model the unknown phase offsets between different observing systems.
        %
%        \begin{figure}
%            \centering
%            \begin{subfigure}{0.97\columnwidth}
%                \centering
%                \includegraphics[width=\columnwidth]{images/mockpulsar1.pdf}
%                \caption{Mock residuals for pulsar J$0437$-$4715$ used in the test of Sec.\ \ref{sec:simspectrum}. We generated 1,500 TOAs over a time span of 6,000 days, injecting power-law timing noise [Eq.\ \eqref{eq:powerlaw}] with parameters $A=3\times10^{-14}$, and $\gamma=4.33$.}
%                \label{fig:mockpsr1res}
%            \end{subfigure}
%            %
%            \begin{subfigure}{0.97\columnwidth}
%                \centering
%                \includegraphics[width=\columnwidth]{images/mockpulsar1-spectrum.pdf}
%                \caption{Recovered Fourier-mode variances for the Metropolis and quasi-Gibbs samplers. The error bars show 1-$\sigma$ standard deviations, and the spectrum of injected noise is shown as the dashed line.}
%                \label{fig:mockpsr1sp}
%            \end{subfigure}
%            \caption{Mock data and spectral estimation in the test of the quasi-Gibbs scheme.}
%        \end{figure}
        \begin{figure*}
            \includegraphics[width=0.8\textwidth]{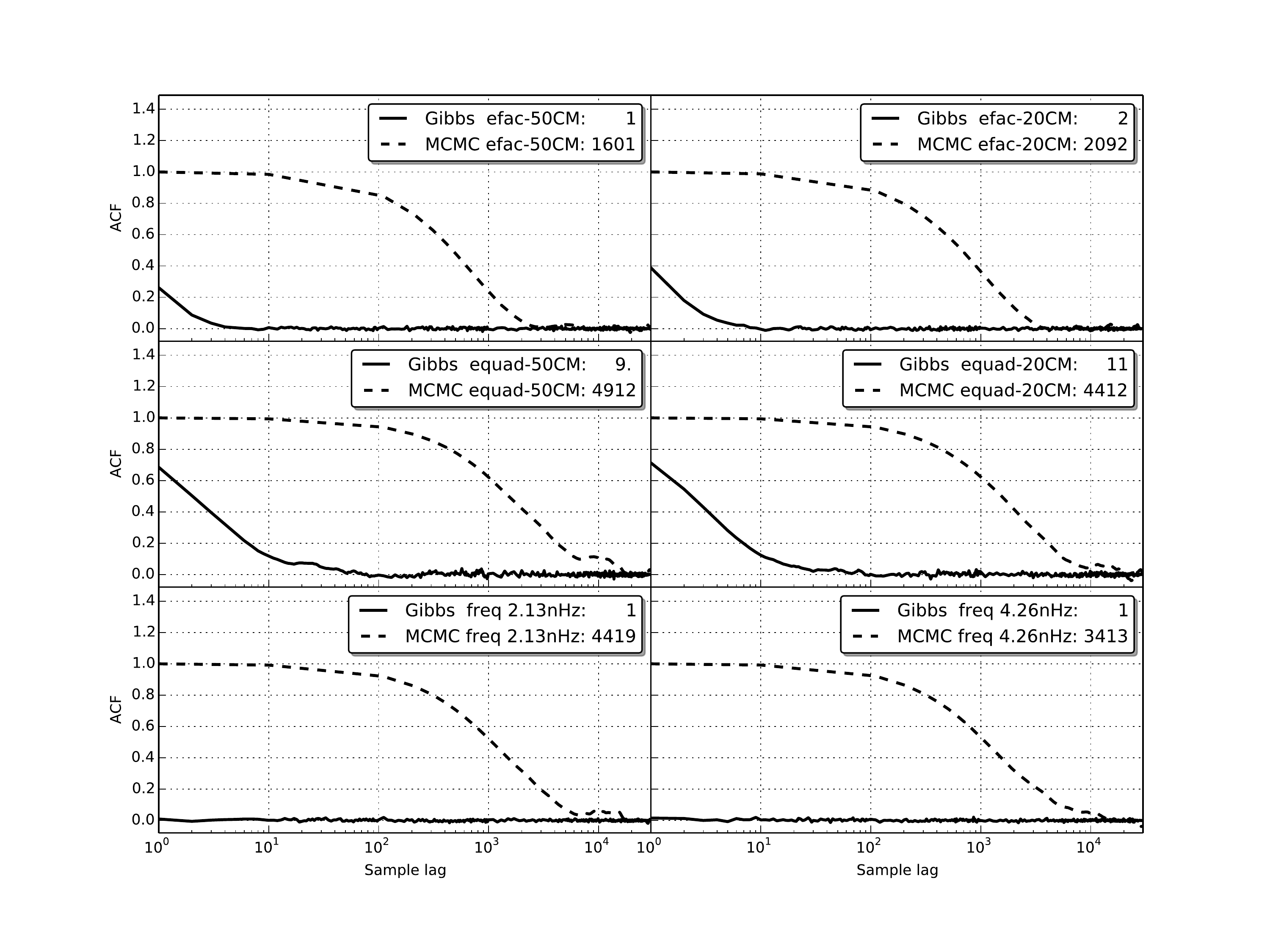}
            \caption{Autocorrelation as a function of sample lag for various model parameters (EFAC and EQUAD for 50- and 20-cm receivers, and Fourier-mode variances at $2.13$ and $4.26$ nHz), as measured in the adaptive-MCMC chain (dashed) and the quasi-Gibbs chain (solid), both run on the mock J$0437$-$4715$ dataset. The legends show the autocorrelation lengths, which were typically 400--1,000 shorter with the quasi-Gibbs sampler.}
            \label{fig:mockpsr1acf}
        \end{figure*}

		The Metropolis sampler was run for 4 million steps, and the quasi-Gibbs scheme for 30,000. The resulting estimates of power spectral density, shown in Fig.\ \ref{fig:mockpsr1res} (bottom panel), agree very well. The autocorrelation functions, shown in Fig.\ \ref{fig:mockpsr1acf}, differ greatly, with much shorter autocorrelation lengths in the quasi-Gibbs scheme---that is why we needed only 30,000 steps for it. 
		The observant reader will note that the autocorrelation length of the Fourier-mode variances is $1$ in the quasi-Gibbs scheme. This is the lowest possible, indicating that our samples are virtually independent draws from the posterior; no sampler can do better. In addition, we note that since the Fourier-mode variances are inherently uncorrelated, their autocorrelation length does not depend on the number of frequencies included in the model, in sharp contrast to Metropolis samplers.

    \subsection{Test of collapsed quasi-Gibbs scheme} \label{sec:simgwb}
        For this test we use mock data for an entire PTA: specifically, the second ``open'' dataset in the IPTA Mock Data Challenge \cite{iptamockdata}, which consists of white radiometer (EFAC) noise in 36 pulsars (with different EFACs) plus a coherently injected GW background with $h_c(1 \, \rm{yr}^{-1}) = 5 \times 10^{-14}$ and $\gamma=4.33$.
        
        In our test we determine the EFACs and the level and shape of the GW background using two sampling schemes: again the ``vanilla'' adaptive Metropolis MCMC of Ref.\ \cite{2014arXiv1404.1267A}, and the collapsed quasi-Gibbs sampler of Sec.\ \ref{sec:modifiedgibbs}. We assume that the GW-background covariance matrix is characterized well by a low-rank expansion that includes $30$ frequency components. The MCMC scheme, which uses a fully marginalized posterior, explores a 38-dimensional parameter space (36 EFACs plus the GW-background $A$ and $\gamma$), while the quasi-Gibbs scheme must deal with a multitude of extra parameters: $2 \times 36 \times 30 = 2,160$ frequency modes, and 36 $\times$ (an average of 12) $= 441$ timing-model parameters, for a whopping total of 2,639.         
        \begin{figure*}
            \includegraphics[width=0.8\textwidth]{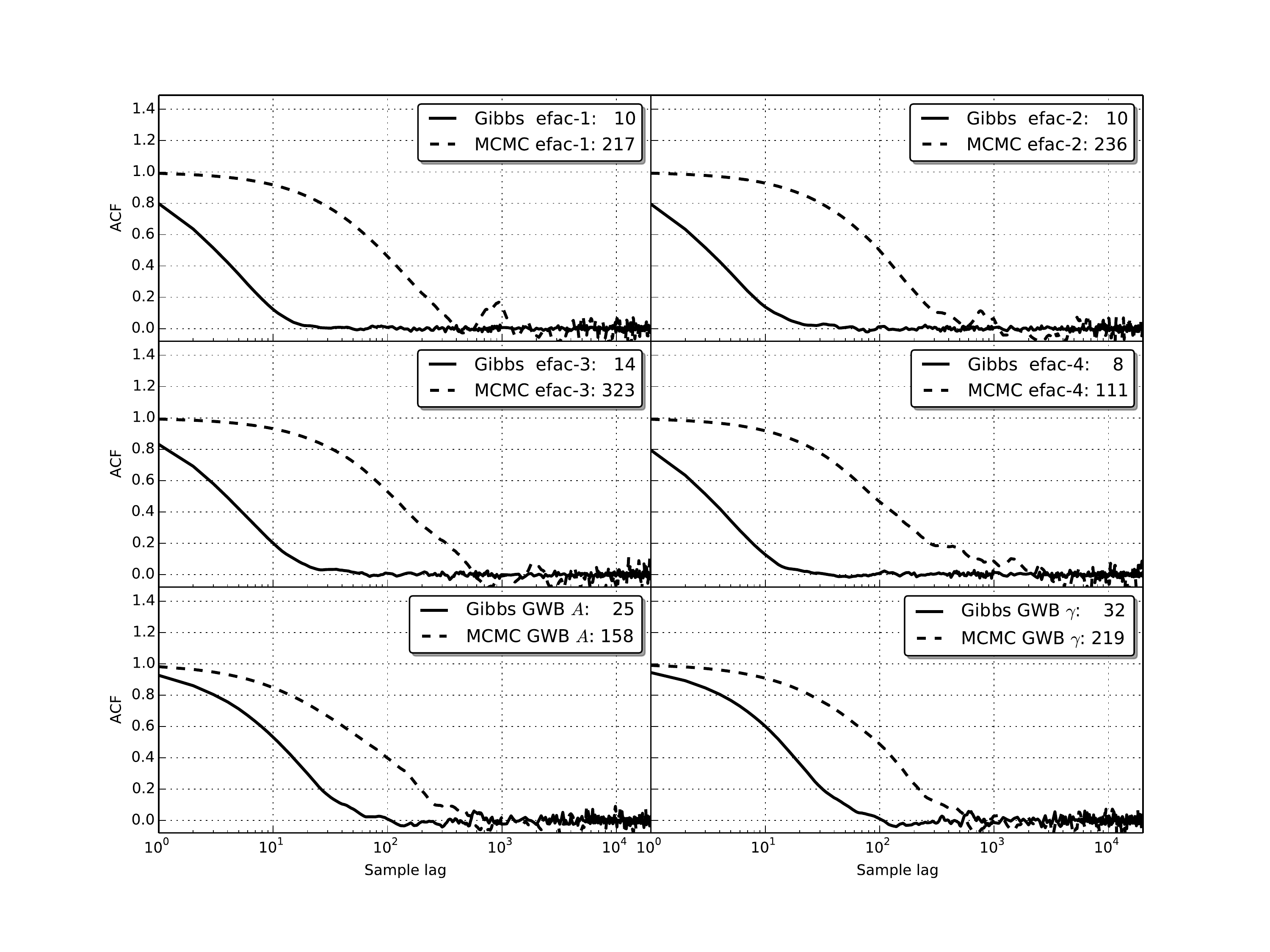}
            \caption{Similar to Fig.\ \ref{fig:mockpsr1acf}, for the adaptive-MCMC chain (dashed) and collapsed quasi-Gibbs chain (solid), both run on the multipulsar dataset form the first IPTA Mock Data Challenge. We plot the autocorrelation functions for four EFAC parameters, and for the two GWB parameters; the legends show the autocorrelation lengths, which are always shorter for the Gibbs sampler, although not as dramatically as in Fig.\ \ref{fig:mockpsr1acf}.}                
            \label{fig:mockptaacf}
        \end{figure*}

        The autocorrelation functions of the two MCMC chains are shown in Fig.\ \ref{fig:mockptaacf}. As it was the case in the first test, the quasi-Gibbs scheme vastly outperforms the adaptive Metropolis MCMC, although it must contend with a much larger parameter space. The smaller autocorrelation lengths result from the fact that the Metropolis--Hastings updates are never performed on all the parameters at once. In fact, the GW-background steps are two-dimensional, and the noise steps are one-dimensional.

        In a more realistic analysis we would have to model also correlated spin noise for every pulsar. The corresponding hyperparameters are highly covariant with those of the GW background, and together they would create a 74-dimensional covariant block, a very significant increase.   
However, that is as bad as it gets: all the other parameters (such as white-noise, jitter-like--noise, and DM-variation hyperparameters) would not further increase autocorrelation lengths. Since 74 covariant dimensions are manageable with modern computing systems, our scheme makes a full-IPTA-sized, full-parameter-set analysis feasible.

\section{Conclusions} \label{sec:conclusions}

	In this paper we have reviewed the description of stochastic signals in pulsar-timing data analysis, which we have recast in the language of Gaussian processes. In this formal context we have rederived and optimized various expressions that are used in Bayesian inference. For some, the Gaussian-process description offers a more insightful interpretation; for others, it allows computationally more efficient implementations.
    
    The Bayesian-inference schemes in current use have trouble scaling up to large datasets such as those assembled by the International Pulsar Timing Array. Their analysis should include full pulsar noise models as in \citet{2014arXiv1404.1267A}, resulting in a very large parameter space to explore. Even with the optimized likelihood expressions that we reviewed in this manuscript, the ensuing MCMC autocorrelation lengths are so large that practical analysis becomes computationally challenging. In this paper we have addressed this problem by constructing two sampling schemes inspired by Gibbs sampling.
    
    The first scheme is very well suited to power-spectral-density estimation in single-pulsar datasets, where we parametrize the power spectrum by independent variance parameters at frequencies multiples of $1/T$, with $T$ the length of the dataset. Currently this is done in practice for few frequencies (up to $\sim 20$ \cite{2013PhRvD..87j4021L}). However, an extended analysis should include many more Fourier modes, possibly all the way up to the Nyquist frequency. In our scheme we partition parameter space in several blocks; for some of them we can draw samples directly from the conditional block posterior; others allow very rapid conditional-posterior evaluations. The parameters in different blocks are almost uncorrelated, resulting in greatly reduced chain autocorrelation lengths. With tests on mock data we demonstrated that the autocorrelation lengths obtained with our Gibbs-inspired sampler are nearly optimal for all Fourier-sum variances, which makes extended spectral analysis practical for single pulsars.

	The second scheme, which we named a collapsed quasi-Gibbs sampler, is well-suited for the Bayesian analysis of full IPTA-sized multipulsar datasets. Unlike the first scheme, this sampler does rely on perturbative Metropolis--Hastings updates, so autocorrelation lengths cannot be minimal. However, by combining blocked updates with the direct sampling of quadratic parameters from their conditional posteriors, we were still able to reduce autocorrelation lengths significantly compared to more conventional MCMC methods. Furthermore, in our Gibbs-like scheme the autocorrelation lengths are much less dependent on the number of noise parameters. This makes full noise modeling in Bayesian methods practical in IPTA-sized datasets: we look forward to actually tackling them in their full glory.

\paragraph*{Acknowledgments.}
We are grateful to many NANOGrav and EPTA colleagues for helpful discussions, and to Scott Ransom for use of the NRAO Nimrod cluster, where our tests were run. RvH is supported by NASA Einstein Fellowship grant PF3-140116.
MV was supported by the Jet Propulsion Laboratory RTD program. The research was carried out at the Jet Propulsion Laboratory, California Institute of Technology, under a contract with the National Aeronautics and Space Administration. Copyright 2014 California Institute of Technology. Government sponsorship acknowledged.

\appendix

\section{Equivalence of the $M$- and $G$-matrix formulations}
\label{sec:mgeq}

The equivalence of Eqs.\ \eqref{eq:mmarglike} and \eqref{eq:gmarglike} (i.e., Eq.\ (18) of Ref.\ \cite{2009MNRAS.395.1005V} and Eq.\ (15) of Ref.\ \cite{2013MNRAS.428.1147V}) is established by the following derivation, which is implied but not shown in Ref.\ \cite{2013MNRAS.428.1147V}.
Consider the full SVD decomposition $M = U \Sigma V^*$ [with dimensions $(n \times n) \times (n \times p) \times (p \times p)$], which is equivalent to the reduced decomposition $F \hat{\Sigma} V^*$ [with dimensions $(n \times p) \times (p \times p) \times (p \times p)$], where $U = [F\,G]$. In particular, the $p$ columns of $F$ span the range of $M$, while the $n - p$ columns of $G$ form the orthonormal completion of $F$ to a full $n$-dimensional basis.

We first concentrate on the determinants that appear at the denominator of Eq.\ \eqref{eq:mmarglike}, obtaining
\begin{widetext}
\begin{equation}
|M^T C^{-1} M| = |V \hat{\Sigma} F^T C^{-1} F \hat{\Sigma} V^*| =
|\hat{\Sigma} \, F^T C^{-1} F \, \hat{\Sigma}| = 
|\hat{\Sigma}|^2 |F^T C^{-1} F|
\end{equation}
(since orthogonal transformations leave determinants invariant, and the determinant of the product of square matrices is the product of their determinants); and
\begin{equation}
|C| = |U^T C U| = |G^T C G| |(F^T C^{-1} F)^{-1}| = |G^T C G| / |F^T C^{-1} F|,
\end{equation}
where the second equality can be read off from the block matrix identity
\begin{equation}
\begin{aligned}
U^T C U = 
\left(\begin{array}{cc}
G^T C G & G^T C F \\
F^T C G & F^T C F 
\end{array}\right) &=
\left(\begin{array}{cc}
G^T C G & 0 \\
F^T C G & I 
\end{array}\right) 
\left(\begin{array}{cc}
I & (G^T C G)^{-1} G^T C F \\
0 & F^T C F - F^T C G(G^T C G)^{-1} G^T C F
\end{array}\right) = \\ 
&=
\left(\begin{array}{cc}
G^T C G & 0 \\
F^T C G & I 
\end{array}\right) 
\left(\begin{array}{cc}
I & (G^T C G)^{-1} G^T C F \\
0 & (F^T C^{-1} F)^{-1}
\end{array}\right).
\end{aligned}
\end{equation}
Thus the normalization factor of Eq.\ \eqref{eq:mmarglike} is given by $\sqrt{(2 \pi)^n |\hat{\Sigma}|^2 |G^T C G|}$; we may drop $|\hat{\Sigma}|$, which is essentially arbitrary (it is the Jacobian of the coordinate transformation $\eta^\prime = (F^TF)^{-1}F^TM\eta$, while we take infinitely vague priors for these parameters), and adjust the $2 \pi$ exponent to match the $n - p$ dimension of $G^T C G$.

Moving on to the main quadratic expression in Eq.\ \eqref{eq:mmarglike},
we rewrite $C' = C^{-1} - C^{-1} M (M^T C^{-1} M)^{-1} M^T C^{-1}$ as
\begin{equation}
\label{eq:demo1}
\begin{aligned}
C' &= 
C^{-1} - C^{-1} (F \hat{\Sigma} V^*) (V \hat{\Sigma} F^T C^{-1} F \hat{\Sigma} V^*)^{-1} (V \hat{\Sigma} F^T) C^{-1} \\
&=
C^{-1} - C^{-1} F (F^T C^{-1} F)^{-1} F^T C^{-1}
\end{aligned}
\end{equation}
where we have used the fact that for unitary $V$ and invertible $X$, $(V X V^*)^{-1} = V X^{-1} V^*$, and that for diagonal $\hat{\Sigma}$ and invertible $Y$, $(\hat{\Sigma} Y \hat{\Sigma})^{-1} = \hat{\Sigma}^{-1} Y^{-1} \hat{\Sigma}^{-1}$. Now, if we apply $C'$ to the data $y$ rewritten as $(F F^T + G G^T) y$, we see that the terms that involve $F^T y$ (either on the right or the left) vanish trivially. For instance,
\begin{equation}
C' (F F^T y) = 
(C^{-1} F - C^{-1} F (F^T C^{-1} F)^{-1} F^T C^{-1} F) (F^T y) =
(C^{-1} F - C^{-1} F) (F^T y) = 0.
\end{equation} 
We are then left with
\begin{equation}
\begin{aligned}
y^T C' y & = (G^T y)^T G^T C' G (G^T y) =
(G^T y)^T (G^T C^{-1} G - G^T C^{-1} F (F^T C^{-1} F)^{-1} F^T C^{-1} G) (G^T y) \\
&= (G^T y)^T (G^T C G)^{-1} (G^T y),
\end{aligned}
\label{eq:laststep}
\end{equation}
\end{widetext}
where the last equality can be proved by direct matrix multiplication.
We thus recover Eq.\ \eqref{eq:gmarglike}.

\section{Data compression}
\label{sec:datacompression}

As we have seen in Sec.\ \ref{sec:lowrank}, we have focused our efforts on
overcoming the bottleneck in evaluating the likelihood on low-rank expansions of
the covariance matrix. We observed that the covariance matrix is the sum
of a diagonal matrix and a rank-reduced matrix, and we applied the Woodbury
lemma in various ways, thereby accurately approximating the likelihood function.

Another approach that utilizes the rank deficiency of various components in the
covariance matrix was formulated by van Haasteren \cite{2013MNRAS.429...55V}, who observed that one is usually not interested in all the parameters
$\theta^\text{(non-TS)}$ in the likelihood function, which allows for the
likelihood function to be modified in a way that retains sensitivity only to the parameters of
interest. This was presented in the form of linear data compression
$\hat{y} = Hy$, with the compression matrix $H$ constructed in a way to maximize
sensitivity to some subset of $\theta^\text{non-TS}$ with its number of columns
as low as possible. In the language of this paper, it means that the information
about our signal of interest is encoded in a small subset of $\phi_\mu$ functions
of the Gaussian process. By using a data vector of reduced size, the
transformed covariance matrix is reduced in size as well, which in turn reduces the
computational burden. Here we present these ideas in a slightly altered way to
conform to the formalism presented in this work.

Essentially, to evaluate the likelihood, we want to approximate two quantities.
$y^{T} C^{-1} y$ and $\det C$ [or when including the timing model, these same
quantities with the $G$-matrix inserted as in Eq.\ \eqref{eq:gmarglike}]. For some
combinations of $D$ and $U$, it is possible to use the approximation
\begin{widetext}
\begin{equation}
    y^{T} C^{-1} y = y^{T} \left(D + U S U^{T}\right)^{-1} \approx y^{T}H
    \left(H^{T}DH + H^{T}USU^{T}H\right)^{-1}H^{T}y + y^{T}H_{c}
    \left(H_{c}^{T}DH_{c}\right)^{-1}H_{c}y.
\label{eq:comprappr}
\end{equation}
\end{widetext}
Here $H$ and $H_c$ are matrices with the properties $H_c^{T} H = 0$, and $HH^{T}
+ H_{c}H_{c}^{T} = I$, and they must be constructed for a specific
problem. It is only possible to find suitable $H$ and $H_c$ when the following
requirements can be satisfied:
\begin{equation}
    H_c^{T} U = 0, \quad HH^{T} U = U, \quad H_c^{T} D H = 0
\label{eq:compreq}
\end{equation}
The authors found that these criteria are sufficiently satisfied only in limited
cases, mainly when the columns of $U$ consist of a basis of Fourier modes as described in Sec.\
\ref{sec:fourier}.
% [RvH: comment on this more. Perhaps some examples].
The matrix $H$ can be constructed from $U$ analogous to how
the $G$-matrix was constructed from $M$ in Sec.\ \ref{sec:timingmodel} with an
SVD. Including the marginalization over the timing model with the $G$-matrix
formalism, we end up with $(H,H_c) = W$, with $W$ from the SVD $W \Sigma V^{*} =
G^{T} U U^{T} G$. Here $H$ consists of the first $l$ columns of $W$, with $l$
the number of non-singular values in $\Sigma$.\footnote{This is typically the
number of columns in $U$, but it can be smaller (numerically) when the timing
basis is sufficiently close to the basis in $U$.}

With Eq.\ \eqref{eq:comprappr} we have made the likelihood function separable,
with one piece greatly rank reduced, and the other part large but with a
diagonal covariance matrix. The bottleneck will be the $O(l^{3})$ inversion of
$H^{T}CH$, or the $O(ln^2)$ operation of the multiplication $H^{T}C$.

Our presentation of data compression differs from the ``ABC method'' originally presented by van Haasteren \cite{2013MNRAS.429...55V}, which did not include both terms of the
separated likelihood function. By only using the data $H^{T}y$, and not
$H_{c}^{T}y$, the ABC method loses sensitivity to some model parameters, and the
actual value of the likelihood is changed. Bayesian model selection is not
possible in that case, or when $H^{T}_{c}DH\ne 0$. We do note that, even when
our likelihood function is not fully separable, Eq.\ \eqref{eq:comprappr}
represents a fully valid way to do analyze the observations. It is equivalent to
partitioning the data in two separate components, and analyzing the components
simultaneously. Some correlation information may have gone lost, but the result
is still internally consistent for \emph{any} $H$. This does not mean that the
parameter estimates are the same for any $H$. Since the data is changed, the
actual estimates can vary.

\bibliography{gp}

\end{document}